\documentclass[pdftex,twocolumn,epjc3]{svjour3}          

\RequirePackage[T1]{fontenc}

\smartqed  

\RequirePackage{graphicx}
\RequirePackage{mathptmx}      
\RequirePackage{flushend}
\RequirePackage[numbers,sort&compress]{natbib}
\RequirePackage[colorlinks,citecolor=blue,urlcolor=blue,linkcolor=blue]{hyperref}

\journalname{Eur. Phys. J. C}

\usepackage{amsmath}
\usepackage{amssymb}
\usepackage{amsfonts}
\usepackage{bbold}
\usepackage{bm}
\usepackage{color}
\usepackage{graphicx}
\usepackage{mathrsfs}
\usepackage{soul}

\usepackage{url}
\def\urlprefix{}
\def\url#1{}

\usepackage{esint}

\usepackage{braket}

\begin{document}

\author{R. Leme\thanksref{e1,addr1,addr2}
        \and
        O. Oliveira\thanksref{e2,addr3}
        \and
        G. Krein\thanksref{e3,addr1}
}

\thankstext{e1}{e-mail: rrleme@ift.unesp.br}
\thankstext{e2}{e-mail: orlando@teor.fis.uc.pt}
\thankstext{e3}{e-mail: gkrein@ift.unesp.br}

\institute{Instituto de F\'{\i}sica Te\'orica, Universidade Estadual Paulista, Rua
Dr. Bento Teobaldo Ferraz, 271 - Bloco II, 01140-070 S\~ao Paulo, Brazil\label{addr1}
          \and
          Universidade Tecnol\'ogica Federal do Paran\'a (UTFPR-CP), 
Avenida Alberto Carazzai 1640, Corn\'elio Proc\'opio, 86300-000 Paran\'a, Brazil\label{addr2}
          \and
          CFisUC, Department of Physics, University of Coimbra, P-3004 516 Coimbra, Portugal\label{addr3}          
}

\date{ }

\title{Approximate dual representation for Yang-Mills SU(2) gauge theory}

\maketitle

\begin{abstract}
An approximate dual representation for non-\-Abelian lattice gauge theories in terms 
of a new set of dynamical variables, the plaquette occupation numbers (PONs) that 
are natural numbers, is discussed. They are the expansion indices of the local 
series of the expansion of the Boltzmann factors for every plaquette of the 
Yang-Mills action. After studying the constraints due to gauge symmetry, the 
SU(2) gauge theory is solved using Monte Carlo simulations. For a PONs 
configuration the weight factor is given by Haar-measure integrals over all links 
whose integrands are products of powers of plaquettes.  Herein, updates are 
limited to changes of the PON at a plaquette or all PONs on a coordinate plane. 
The Markov chain transition probabilities are computed employing truncated maximal 
trees and the Metropolis algorithm. The algorithm performance is investigated with 
different types of updates for the plaquette mean value over a large range of $\beta$s. 
Using a $12^4$ lattice very good agreement with a conventional heath bath algorithm 
is found for the strong and weak coupling limits. Deviations from the latter being 
below 0.1\% for $2.5 < \beta < 3$. The mass of the lightest $J^{PC}=0^{++}$ glueball 
is evaluated and reproduces the results found in the literature.
\end{abstract}

%
\section{Introduction}

The computation of the properties of strongly-interacting matter directly from Quantum 
Chromodynamics (QCD) remains a challenging problem. For matter at zero baryon density, Monte Carlo 
lattice QCD simulations are currently used to address both zero and finite temperature~\cite{gattBook}. 
On the other hand, the investigation of dense quark matter, as required for example 
to study the structure of atomic nuclei and neutron stars, the quark-gluon plasma produced 
in heavy ion collisions, and the matter that existed in the early stages of the Universe, 
is still an open problem for lattice QCD simulations due to algorithmic limitations.
Indeed, the investigation of such systems, e.g. in the grand canonical ensemble, demands the 
introduction of a finite chemical potential $\mu$ in the partition function of the theory. 
The baryon chemical potential turns the Euclidean action into a complex-valued function
and the integration measure in the path integral of the partition function is no longer 
positive definite, giving rise to the so-called sign problems, and thereby limiting the use of Monte 
Carlo techniques with importance sampling. For sufficiently small values of $\mu$, the study of dense 
systems can still rely on importance sampling when combined with 
re-weighting~\cite{Splittorff:2007mr,Splittorff:2007ck}. However, in general, the handling of 
complex actions requires the introduction of new sampling techniques as, for example, the direct 
sampling of the density of states or a mapping of the theory into new variables such that one 
recovers a positive Boltzmann factor; in this latter approach, the theory reformulated in terms of 
the new variables is called the dual theory {\textemdash} Ref.~\citep{Gattringer:2016kco} provides
a recent review on these methods for lattice field theories. The mapping of a given theory into its 
dual has been used to overcome sign problems appearing in different fields~\citep{Aarts:2013naa}.

In lattice QCD in the strong coupling limit, sign problems can be avoided by mapping the theory 
into a dual representation, using new ``dual variables'', after the integration of the gauge fields 
prior to the integration of the fermion fields~\citep{deForcrand:2014tha,deForcrand:2009dh,Rossi:1984cv}. 
The gauge symmetry of the original theory imposes constraints on the new set of dual variables 
which, nevertheless, can be handled via generalizations of the original  Prokof'ev-Svistunov worm 
algorithm~\citep{Prokofev:2001ddj}.

Another example of a dual representation of QCD is the effective theory 
introduced in Ref.~\citep{DeGrand:1983fk}, where the fundamental degrees of freedom are
the Polyakov loops defined in the group $\mathbb{Z}(3)$. This effective theory can be derived from 
QCD in the strong coupling limit, by restricting the non-Abelian gauge degrees of freedom to the center 
of the group SU(3), i.e. to the group $\mathbb{Z}(3)$, and performing a hopping expansion in the quark 
sector. The action of the $\mathbb{Z}(3)$ effective theory inherits a sign problem from QCD. 
However, after rewriting the original partition function in terms of dual variables, it becomes a sum of 
real and positive Boltzmann weights~\citep{Mercado:2011ua}. The dual variables are {\em dimers}, that are 
attached to the lattice links, and {\em monomers}, that are attached to the lattice sites. In the dual
representation, the complex nature of the original action is washed out~\citep{Mercado:2011ua}. 
Symmetries of the original theory appear, again, as constraints on the dual variables of the 
reformulated theory that can be handled with the use of a generalized worm algorithm. 

In recent years several other interesting QCD-related theories were studied using dual representations. 
Theories with O(N) and CP(N-1) symmetries, which, like QCD, are asymptotically free, were investigated
with dual representations at zero~\citep{Wolff:2009kp,Wolff:2010qz} and finite density~\citep{Bruckmann:2015sua}. 
Strongly interacting fermionic theories, relevant for graphene~\citep{CastroNeto:2009zz} and also for particle
physics, were investigated with the fermion bag approach~\citep{Ayyar:2014eua,Chandrasekharan:2013rpa,
Chandrasekharan:2013aya,Chandrasekharan:2011mn}, in which a dual representation can be built after a suitable
integration of the fermionic degrees of freedom. By combining strong coupling and hopping-parameter expansions, 
an effective theory~\citep{Fromm:2011qi} in the dual representation free of sign problems is obtained. Scalar 
field theories have also been successfully mapped into dual representations, see e.g. 
Refs.~\citep{Gattringer:2012df,Korzec:2011gh,Gattringer:2012ap,Endres:2006xu,Rindlisbacher:2016zht}.

In what concerns gauge field theories, dual representations were implemented for pure Abelian 
U(1) theory~\citep{Panero:2005iu,Azcoiti:2009md}, Higgs-U(1) theory~\citep{Mercado:2013ola,Mercado:2013yta}, 
and U(1) Abelian theory with fermion fields~\citep{Adams:2003cca}. For pure SU(N) lattice gauge theory a
dual representation was suggested recently in Ref.~\citep{vairinhos}, where the dual variables are 
random Gaussian matrices introduced by recursive applications of the Hubbard-Stratonovich 
transformation~\citep{hubbard}.  Recently, a dual representation for non-Abel\-ian gauge theories 
was suggested in Refs.~\citep{{Marchis:2017oqi},Gattringer:2011gq,Marchis:2016cpe}, in which the partition 
function for the dual theory is given by a sum of positive and negative terms, which prevents 
the use of Monte Carlo simulations with importance sampling to solve the theory.

In the present work we discuss a new approximate dual representation for pure 
non-Abelian gauge theories. Starting from the partition function written in terms of the Wilson 
action, we expand the Boltzmann exponential factor of a single plaquette as a power series. The 
expansion indices of each plaquette, $b_{\mu\nu}\left(x\right)\in\mathbb{N}_{0}$, where
$\mathbb{N}_{0}$ is the set of natural numbers, after integrating over the gauge fields, play 
the role of dynamical variables. The Boltzmann weights become functions of the dual variables 
$b_{\mu\nu}\left(x\right)$, and a Metropolis-type algorithm can be built. The transition probabilities 
of the corresponding Markov chain are ratios between these weights. In the new representation of the 
non-Abelian gauge theory, the weights are computed using the Haar-measure integrals involving the link 
variables. The integration over the links is a non-trivial problem \textit{per se} as each link 
is coupled to all links in the entire lattice. For the numerical experiment, we make approximations 
in the integration over the link fields to estimate the transition probability defining the Markov chain 
and thus generate ensembles of the dual variables $\left\{ b_{\mu\nu}\left(x\right)\right\} $.

The work reported herein investigates the pure SU(2) Yang-Mills gauge theory. Although the boson sector 
of a gauge theory does not suffer from the sign problem, our goal is to test a new algorithm/representation 
of a gauge theory to study strong interactions. The natural development of the ideas discussed herein are 
both the inclusion of matter fields to simulate the full theory and the improvement in the approximations
considered. The rationale used here to build a new dual representation can, in principle, be extended 
to the fermionic sector. The full theory requires the use of an enlarged set of dynamical variables, 
defined after the expansion of the partition function. Furthermore, the constraints in the corresponding 
dual theory due to the gauge symmetry are of the same type as those for the pure Yang-Mills theory. 
On the other hand, the integration over the link fields requires a new analysis.

We test our algorithm by computing the plaquette mean value, related to the energy 
density of the pure gauge system, over a large range of the lattice coupling constant $\beta$ 
and the mass of the (expected) lightest scalar glueball state ($J^{PC}=0^{++}$).
Our results show that the plaquette mean value obtained with the algorithm developed here deviates, 
in the worst case, by less than $0.1$\% when compared with a standard heat bath simulation for 
$\beta\in[0, \, 4.5]$. The mass of the lightest glueball agrees well with previous lattice estimates 
\citep{Chen:2005mg,Lucini:2004my,Albanese:1987ds,Teper:1987wt,Berg:1982kp,Falcioni:1982ja,Berg:1980gz}
and also with estimates based on a gauge-gravity duality model~\citep{Brunner:2015yha}.

Our paper is organized as follows. In the next section we present
our approximate dual representation for the non-Abelian Yang-Mills theory. 
In Sec.~\ref{sub:Constraint-over-the} we discuss the constraints on the dual variables due to
gauge symmetry which determine the types of updates that must be considered
in a algorithm approach to solve the theory.
In Sec.~\ref{sec:Algoritmo} we discuss the Monte Carlo algorithm used in our approach.
We also present strategies to decouple a region from the entire lattice surrounding a 
dual variable to be updated locally. In the factorized region, the group integrals are done analytically. 
Still in Sec.~\ref{sec:Algoritmo} we show how to implement one possible type of nonlocal update. 
In Sec.~\ref{sec:Observables} we show how to represent the observables to be measured  in terms of the dual variables.
In Sec.~\ref{sec:Monte-Carlo-results} we give the details of the simulations and
report the numerical data for the observables measured.
A summary in Sec.~\ref{sec:Conclusions} completes the paper.

\section{\label{sec:Lattice-gauge-system}Approximate dual representation for lattice 
Yang-Mills theory}

The lattice formulation of pure Yang-Mills theory uses as fundamental
fields the link variables $U_{\mu}(x)$, which belong to the gauge
group SU(N). We consider the standard Wilson action~\cite{Wilson:1974sk}: 
\begin{eqnarray}
S\left[U\right]=\frac{\beta}{N}\sum_{x\in V}\sum_{\mu<\nu}
\textrm{Re Tr}\left[\mathbb{1}-U_{\mu\nu}(x)\right],
\label{wilsonAction}
\end{eqnarray}
with the plaquette $U_{\mu\nu}(x)$ given by the product of link variables
\begin{equation}
U_{\mu\nu}(x) = U_{\mu}(x)U_{\nu}(x+\hat{\mu})U_{\mu}^{\dagger}(x+\hat{\nu})U_{\nu}^{\dagger}(x),
\end{equation}
where the spacetime indices $\mu$ and $\nu$ run from 1 to $d$, with
$d$ being the dimension of the Euclidean space, and $x$ runs over the
lattice volume $V$. The partition function of the theory is given by 
\begin{equation}
Z  =  C \int\mathscr{D}U 
\prod_{x, \mu<\nu} e^{\frac{\beta}{N} \, \textrm{Re Tr}\left[U_{\mu\nu}(x)\right]},
\label{eq:partition_func}
\end{equation}
where
$\mathscr{D}U=\prod_{x,\mu}dU_{\mu}(x)$ is the Haar measure for the gauge links,
$C= \exp \left(-\beta N^{V_{p}-1}\right)$ is a normalization factor
and $V_{p}$ is the number of plaquettes in the volume $V$.
Given an operator $\mathcal{O}(U)$,
its vacuum expectation value is represented by the functional integral
\begin{equation}
\langle\mathcal{O}\rangle = \frac{C}{Z} \int\mathscr{D}U~\prod_{x,\mu<\nu} 
e^{\frac{\beta}{N}\,\textrm{Re Tr}\left[U_{\mu\nu}(x)\right]}~\mathcal{O}(U).
\label{eq:partition_func_O}
\end{equation}
In the traditional lattice approach, this expectation value is estimated by the average
\begin{equation}
\langle\mathcal{O}\rangle\approx\frac{1}{N_{conf}}\sum_{i=1}^{N_{conf}}\mathcal{O}(U^{(i)}),
\end{equation}
where the set of configurations $\mathscr{U}=\{U^{(i)}, i=1, \cdots, N_{conf}\}$, distributed 
according to $\exp\{-S[U]\}$, is produced with a Monte Carlo algorithm. The statistical error 
associated with such an estimate scales with the number of configurations as $N_{conf}^{-1/2}$.

The simulation of Yang-Mills theory with a dual representation demands rewriting
the partition function in Eq.~(\ref{eq:partition_func}) in terms of a new
set of dynamical variables other than the links. In order to be able
to apply such a type of algorithm, let us expand the exponentials appearing
in the partition function in powers of $\beta$ 
\begin{eqnarray}
Z = \int\mathscr{D}U\prod_{x, \, \mu \, < \, \nu} ~ ~ \sum_{b_{\mu\nu}(x)}\frac{\left[\frac{\beta}{N}\textrm{Re Tr }U_{\mu\nu}(x)\right]^{b_{\mu\nu}(x)}}{b_{\mu\nu}(x)!} ,
\end{eqnarray}
where we have discarded for the moment the global factor~$C$. Performing
the integration over the link variables, i.e. computing the integral
$\int\mathscr{D}U$, $Z$ can then be viewed as a function of the
discrete set of variables $b_{\mu\nu}(x)$, which are natural numbers.
Let us introduce the notation 
\begin{equation}
\sum_{\left\{ b\right\} }=\prod_{x, \mu<\nu} \sum_{b_{\mu\nu}(x)} ,
\end{equation}
so that the partition function can be written as 
\begin{equation}
Z=\sum_{\left\{ b\right\} }Q_{\left\{ U\right\} }\left[\left\{ b\right\} \right],
\label{eq:partition_Function_New}
\end{equation}
where 
\begin{eqnarray}
Q_{\left\{ U\right\} }\left[\left\{ b\right\} \right] &=& \int\mathscr{D}U
\prod_{x, \mu<\nu}
\frac{\left[\frac{\beta}{N}\textrm{Re Tr }U_{\mu\nu}(x)\right]^{b_{\mu\nu}(x)}}{b_{\mu\nu}(x)!}
.\label{eq:weight_Q}
\end{eqnarray}
The integration of the link variables defines the weight functions 
$Q_{\left\{ U\right\} }\left[\left\{ b\right\} \right]$
which are, themselves, functions of the natural numbers $b_{\mu\nu}\left(x\right)$,
the new dynamical variables; $b_{\mu\nu}\left(x\right)$ are from now on called 
plaquette occupation number (PON). Then, one can define a Markov chain to update 
the $b_{\mu\nu}(x)$ values by choosing a transition probability given by the ratio of
the weight functions $Q_{\left\{ U\right\} }\left[\left\{ b\right\} \right]$,
that complies with the principle of detailed balance and ensures the
convergence of the Markov chain to the right probability distribution.
Before dealing with the details of the update, let us discuss the
constraints on $b_{\mu\nu}(x)$ due to the group integration over
the link variables.

\section{Constraints on the dual variables $b_{\mu\nu}(x)$ 
\label{sub:Constraint-over-the}}

Herein we discuss the constraints on the $b_{\mu\nu}(x)$ when group-integrating over the gauge 
links. The results and the group integrations discussed below can, in principle, be extended to 
SU(N) but we restrict our analysis to SU(2). The main properties and results for the 
group integration required to understand the current work are summarized in the 
\ref{sec:Group-integration}.

\begin{figure}[t]
\centering
\includegraphics{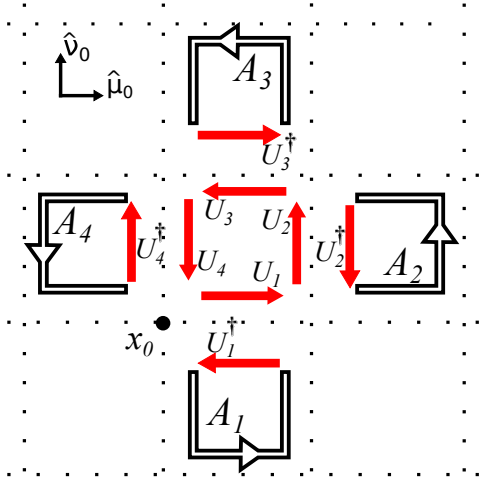}

\caption{Representation of the $(\mu_{0},\nu_{0})$ lattice plane. The gauge links are shown 
as arrows. The ``central plaquette'' ~$U_{\mu_{0}\nu_{0}}(x_{0})= U_{1}U_{2}U_{3}U_{4}$
(in solid red arrows) and the plaquettes which contain any of the links appearing in $U_{\mu_{0}\nu_{0}}(x_{0})$.
$A_{i}$ are the staples (in non-solid black lines), defined in the $(\mu_{0},\nu_{0})$
plane, required to complete the neighboring plaquettes besides the links
in the ``central plaquette''.}
\label{fig:plane} 
\end{figure}


Let us consider the plaquette 
\begin{eqnarray}
U_{\mu_{0}\nu_{0}}(x_{0}) & = & U_{\mu_{0}}(x_{0})U_{\nu_{0}}(x_{0}+\hat{\mu}_{0})
U_{\mu_{0}}^{\dagger}(x_{0}+\hat{\nu}_{0})U_{\nu_{0}}^{\dagger}(x_{0}) \nonumber \\
& = & U_{1}\,U_{2}\,U_{3}\,U_{4} ,
\end{eqnarray}
defined on the $(\mu_{0},\nu_{0})$ plane, see Fig.~\ref{fig:plane}, where $U_{l}$ with $l=1,\cdots,4$ 
stands for a generic link and $l$ is a composite index taking values in the set: 
\begin{equation}
L = \left\{ \left(x_{0},\mu_{0}\right);\left(x_{0}+\hat{\mu}_{0},\nu_{0}\right);\left(x_{0}+\hat{\nu}_{0},\mu_{0}\right);\left(x_{0},\nu_{0}\right)\right\}.
\label{eq:coordenadas_plaquetta}
\end{equation}

Let $A_{i}$ be the staple that together with the link variable $U_{i}$
defines a plaquette in the $(\mu_{0},\nu_{0})$ plane which shares
with $U_{\mu_{0}\nu_{0}}(x_{0})$ the link $U_{i}$ {\textemdash} see Fig.~\ref{fig:plane}. 
In the following, to simplify the notation, we will write $b_{l}$ for the dynamical 
variable that is associated with the plaquette containing the staple $A_{l}$,
i.e. the staple in the plane $(\mu_{0},\nu_{0})$ that is associated
with the link $U_{l}$. The weight function $Q$ associated with the plaquettes represented
in Fig.~\ref{fig:plane} is 
\begin{eqnarray}
Q_{\left\{ U\right\} }\left[\left\{ b\right\} \right] &=& 
\left(\frac{\beta}{N}\right)^{\sum b_{\mu\nu}(x)}
\int\mathscr{D}U\nonumber \\
&& \times \, \frac{1}{b_{0}!}\left[\textrm{Tr }U_{1}U_{2}U_{3}U_{4}\right]{}^{b_{0}}\nonumber \\
&& \times \, \frac{\left[\textrm{Tr }U_{1}A_{1}^{\dagger}\right]^{b_{1}}}{b_{1}!}
\times\frac{\left[\textrm{Tr }U_{2}A_{2}^{\dagger}\right]^{b_{2}}}{b_{2}!}\nonumber \\
&& \times \, \frac{\left[\textrm{Tr }U_{3}A_{3}^{\dagger}\right]^{b_{3}}}{b_{3}!}
\times\frac{\left[\textrm{Tr }U_{4}A_{4}^{\dagger}\right]^{b_{4}}}{b_{4}!}\cdots.
\label{eq:function_Q_su2_2}
\end{eqnarray}
The properties of the group integration are such that most of the possible sets 
$\left\{ b\right\}$ have a null weight
and do not contribute to the partition function.
The non-vanishing contributions are those where a given link variable $U_l$, with $l=(\mu,x)$, 
appears $n_l$ times in the integrand, with $n_l$ being a multiple of N, where N is the number of colors.
Consider, for example, the link variable $U_{2}$ in Eq.~\eqref{eq:function_Q_su2_2}: it appears 
$b_{0}+b_{2}=n_{2}$ times in the integrand, i.e.
\begin{equation}
\int dU_2\,(U_2)_{i_1j_1}\,(U_2)_{i_2j_2}\,\cdots\,(U_2)_{i_{n_{2}}j_{n_{2}}},
\end{equation}
and this gives a non-vanishing contribution to $Q_{\left\{ U\right\} }\left[\left\{ b\right\} \right]$
only if $n_2 = b_{0}+b_{2}$ is a multiple of N. This implies that $b_{0}$ and $b_{2}$ are
either multiples of N or their sum is a multiple of N, despite
$(b_{0}\mod N)\neq0$ and $(b_{2}\mod N)\neq0$. In four dimensions,
the link $U_{2}$ belongs to the plaquettes represented in Fig.~\ref{fig:plane}
and also to plaquettes belonging to orthogonal planes not shown in the figure.
Therefore, for a generic link $U_{\mu}(x)$, it follows that the sum over the set
$\{b_{\mu\nu}(x)\}$ that count the number of times $U_{\mu}(x)$ appears in the integral
in Eq.~(\ref{eq:function_Q_su2_2}) is given by 
\begin{eqnarray}
n_{\mu}(x) &=& \sum_{\nu=1}^{\mu-1}\left[b_{\nu\mu}(x)+b_{\nu\mu}(x-\hat{\nu})\right]
\nonumber \\
&& + \sum_{\nu=\mu+1}^{d} \left[b_{\mu\nu}(x)+b_{\mu\nu}(x-\hat{\nu})\right],
\label{eq:total_flux}
\end{eqnarray}
and only those $\{b_{\mu\nu}(x)\}$ configurations such that all $\{n_{\mu}(x)\}$ 
are multiples of N contribute to the partition function, i.e.\begin{eqnarray}
n_{\mu}(x)\!\!\!\mod{\rm N} = 0.
\label{eq:vinculo}
\end{eqnarray}
This is a non trivial constraint that also simplifies the analysis of the possible 
sets of updates that can appear within a Markov chain.


\section{\label{sec:Algoritmo}Update Algorithm}

A possible local update compatible with Eq.~(\ref{eq:vinculo}) replaces
$b_{\mu\nu}(x)\rightarrow b_{\mu\nu}(x)\pm\Delta$, with $\Delta$
being a multiple of~N. In this way, if the original configuration $\{b_{\mu\nu}(x)\}$
verifies the constraint in Eq.~(\ref{eq:vinculo}), the new configuration
is also compatible with Eq.~(\ref{eq:vinculo}). On the other hand, if
$(\Delta\!\!\mod{\rm N})\neq0$, then to fulfil Eq.~(\ref{eq:vinculo}) at all
lattice points, one has to change the $b$
values in the neighboring points accordingly and, therefore, in the
next neighboring points and so on and so forth. The updates where
$\Delta$ is not an integer multiple of N requires a global update
over a finite region of the lattice.

An ergodic algorithm must access all possible $b$ values and, therefore,
requires the use of both local and nonlocal updates. If, for example,
the Markov chain is initiated setting all PON such that $(b_{\mu\nu}(x)\!\!\mod{\rm N})=0$
and only local updates are implemented, i.e. a given $b$ is modified by adding an
integer multiple of N, configurations where all PONs of a given plane are not 
multiples of N cannot be reached and, therefore, the update does not verify the 
ergodicity requirement.

To ensure convergence to the right probability distribution,
one needs to set a detailed balance equation compatible with Eq.~(\ref{eq:vinculo}).
Our implementation chooses randomly a $b$ or a set of $b$'s and proposes new values $b'$. 
As usual in algorithms of this kind, the transition probability for accepting the new 
$b'$ is given by 
\begin{equation}
p = \frac{P_{\{b\}^{\rm old} \rightarrow \{b\}^{\rm new}}}{P_{\{b\}^{\rm new}
\rightarrow \{ b\}^{\rm old}}}
= \frac{Q\left[\{ b\}^{\rm new}\right]}{Q\left[\{ b \}^{\rm old}\right]},
\label{Eq:TransProb}
\end{equation}
which is enough to ensure that the sampling reproduces the correct
distribution probability \cite{Newman1999}. 
 
The computation of the weight function $Q$ requires integration over the link variables, for
all possible PONs configurations, which \textit{per see} is a difficult problem. The Haar measure 
for the group integration is invariant under gauge transformations and this allows rotating 
the links and, eventually, replace some of them by the identity in the evaluation of the 
$Q[\{b\}]$ functions. In particular,  a path in which the maximal number of links allowed by 
the group integration are rotated to the identity defines what is known as 
a ``maximal tree"~\cite{creutz}.
Our proposal consists in, given a $b_{\mu\nu}(x)$ variable to be updated, 
performing an exact integration of the gauge links in the neighborhood of the
plaquette $U_{\mu\nu}(x)$. In order
to be able to compute the transition probability $p$ we set a small number of
links to the identity matrix and, in this way,
decouple a region with links ``close'' to the $b_{\mu\nu}(x)$ variable to be updated
and a region with the remaining ``distant'' links. The transition probability for accepting
the new value $p$ is given by the ratio between weight functions and, therefore, 
the integration of the ``distant'' links cancels out and we need consider only the contributions 
of the links that are closer to the plaquette associated with $b_{\mu\nu}(x)$.
The replacement of a small subset of link variables by the identity matrix is 
clearly an approximation, but it enables to perform group integrations analytically.

\subsection{Local update \label{Sec:groupinte}}

To illustrate our update scheme, let us start considering 
the crudest approximation possible in the local update of a given plaquette occupation number, say 
$b_{\mu_0 \nu_0}(x_0)$, associated with the central plaquette represented in Fig.~\ref{fig:plane}, i.e.
replace all the staples that are connected with $U_{\nu_{0}}(x_{0})$ (the link $U_4$ in
the figure) by the identity. 
Then, the integration over $U_{\nu_{0}}(x_{0})$ is decoupled from the integrations 
over the remaining links and 
\begin{eqnarray}
Q\left[\left\{ b\right\} \right] &\approx& 
\int dU_{\nu_{0}}(x_{0})\left[\textrm{Tr }U_{\nu_0}(x_{0})\right]^{n_0}
Q'\left[\left\{ b\right\} '\right]
\nonumber \\
&=& Q'\left[\left\{ b\right\} '\right]\int dU_{\nu_{0}}(x_{0})
\left[\textrm{Tr }U_{\nu_0}(x_{0})\right]^{n_0},
\label{Eq:Qfact}
\end{eqnarray}
where $n_0 = n_{\nu_0}(x_0)$ is calculated from Eq.~(\ref{eq:total_flux}), 
and  $Q'\left[\left\{ b\right\} '\right]$ is independent of the link $U_{\nu_{0}}(x_{0})$. 
In this way, the transition probability of the local update of the PON $b_{0}=b_{\mu_{0}\nu_{0}}(x_{0})$ is
\begin{eqnarray}
p &=& \frac{Q\left[\left\{ b\right\}^{\rm new}\right]}{Q\left[\left\{ b\right\}^{\rm old}\right]}
\nonumber \\
& \approx & \frac{Q'\left[\left\{ b'\right\} \right]}{Q'\left[\left\{ b'\right\} \right]}
\frac{\int dU_{\nu_{0}}(x_{0})\left[\textrm{Tr }U_{\nu_{0}}(x_{0})\right]^{n_0({\rm new})}}
{\int dU_{\nu_{0}}(x_{0})\left[\textrm{Tr }U_{\nu_{0}}(x_{0})\right]^{n_0(\rm old)}}\nonumber \\[0.2true cm]
& = & K \frac{\int dU_{\nu_{0}}(x_{0})\left[\textrm{Tr }U_{\nu_{0}}(x_{0})\right]^{n_0(\rm new)}}{\int dU_{\nu_{0}}(x_{0})\left[\textrm{Tr }U_{\nu_{0}}(x_{0})\right]^{n_0({\rm old})}},
\label{eq:approx_1}
\end{eqnarray}
where
\begin{eqnarray}
K \equiv \left( \frac{\beta}{\textrm{N}} \right)^{b_0^{(\rm new)}-b_0^{(\rm old)}} \frac{b_0^{(\rm old)}!}{b_0^{(\rm new)}!}.
\end{eqnarray}

To improve on the estimation of $p$, couplings of $U_{\nu_0}(x_0)$ to neighboring links need to be considered. 
A possible next level of approximation is to set all the staples associated with $U_{\nu_{0}}(x_{0})$ 
to the identity with exception of $g= U_{1}U_{2}U_{3}$ (see Fig.~\ref{fig:plane}), then 
\begin{equation}
Q\left[\left\{ b\right\} \right]\approx Q''\left[\left\{ b\right\} ''\right]~\int dU_{\nu_0}(x_0)\,
F\left[U_{\nu_0}(x_0),g\right],\label{eq:basic_int}
\end{equation}
where 
\begin{equation}
F\left[U_{\nu_{0}}(x_{0}),g\right] = \left[\textrm{Tr }U_{\nu_{0}}^{\dagger}(x_{0})\right]^{n_0-b_{0}}
\left[\textrm{Tr }U_{\nu_{0}}(x_{0}) g\right]^{b_{0}},
\label{eq:orlando1}
\end{equation}
and $Q''\left[\left\{ b\right\} ''\right]$ is the group integral
over all the lattice links except for  $U_{\nu_{0}}(x_{0})$.
Now, since $Q''\left[\left\{ b\right\} ''\right]$ and the integral in Eq.~(\ref{eq:basic_int}) share the 
links $U_{1}$, $U_{2}$, and $U_{3}$, they do not decouple and this would not allow us to
obtain a number for the transition probability $p$.
However, 
as we shall discuss in the next two subsections, one can still devise a strategy that allows 
us to integrate over $U_{\nu_{0}}(x_{0})$ taking into account couplings with neighboring 
links, so that under a local update $b^{\rm old}_{0}\rightarrow b^{\rm new}_{0}$, the transition 
probability is the positive real number given by
\begin{eqnarray}
p = \frac{Q[\left\{ b\right\} ^{{\rm new}}]}{Q[\left\{ b\right\}^{\rm old}]} 
& \approx & \frac{\int\widetilde{\mathscr{D}U}\,
F\left[\mathcal{U},\mathcal{B},b^{\rm new}_0\right]}
{\int\widetilde{\mathscr{D}U}\,F\left[\mathcal{U},\mathcal{B},b^{\rm old}_0\right]} \ ,
\label{eq:Q_ratio_approx}
\end{eqnarray}
where $F$ contains through $\mathcal{U}$ a subset of all links $U_{\mu}\left(x\right)$ of
the lattice that are integrated, and $\mathcal{B}$ stands for the PONs associated with the PON 
$b_{0}$ which is being updated. 

The Monte Carlo updates considered in the present work approximate ratios of weight functions $Q$ following the strategy just discussed. 
The integration of the functions $F$ all give positive definite answers and, thus, the approximate ratio 
between the dual Boltzmann weights $Q$ to estimate the transition probability $p$ is also a positive real number.


\subsubsection{Integration over a short path\label{sub:Simple-integration-update}}

Let us consider Fig.~\ref{fig:plane} and the central plaquette associated
with the dual variable $b_{0}=b_{\mu_{0}\nu_{0}}(x_{0})$. The links belonging to this
plaquette (solid red arrows) also contribute to the staples $A_{i} = \{A_1, A_2, A_3, A_4\}$ (non-solid black lines). Recall 
that the aim is to update $b_{0}$ and compute the transition probability~$p$.

A maximal tree can be built by rotating some, but not all, staples associated
with the links $U_{i}$ in the plaquette $U_{\mu_{0}\nu_{0}}(x_{0})$ to the identity
matrix. However, assuming that all the links in $A_{i}$ can be set
to the identity, the group integration can be factorized and one has
to consider only the following integrating function 
\begin{eqnarray}
F_{4} & = & F_{4}\left[\mathcal{U},\mathcal{B},b_{0}\right]\nonumber \\
& = & \frac{1}{b_{0}!}\left(\frac{\beta}{N}\right)^{b_{0}}
\textrm{Tr}\left[U_{1}U_{2}U_{3}U_{4}\right]^{b_{0}}\nonumber \\
&& \times \, \textrm{Tr}\left[U_{1}^{\dagger}\right]^{c_{1}}
\textrm{Tr}\left[U_{2}^{\dagger}\right]^{c_{2}} \textrm{Tr}\left[U_{3}^{\dagger}\right]^{c_{3}}
\textrm{Tr}\left[U_{4}^{\dagger}\right]^{c_{4}},
\label{Eq:F4}
\end{eqnarray}
with the integration measure given by
\begin{equation}
\widetilde{\mathscr{D}U_{4}} = dU_{1}dU_{2}dU_{3}dU_{4}.
\label{eq:haar41}
\end{equation}
The set $\mathcal{U}=\left\{ U_{1},U_{2},U_{3},U_{4}\right\} $ contains
the link variables to be integrated. The set $\mathcal{B}$ contains
the PONs that couple the plaquette $U_{\mu_{0}\nu_{0}}\left(x_{0}\right)$
with the neighboring plaquettes and in two dimensions 
\begin{eqnarray}
\mathcal{B} & = & \bigl\{ b_{\mu_{0}\nu_{0}}(x_{0}+\hat{\mu}_{0}),
b_{\mu_{0}\nu_{0}}(x_{0}-\hat{\mu}_{0}),b_{\mu_{0}\nu_{0}}(x_{0}+\hat{\nu}_{0}), 
\nonumber \\
&& \, b_{\mu_{0}\nu_{0}}(x_{0}-\hat{\nu}_{0})\bigr\}.
\end{eqnarray}
For a generic dimensionality, the set $\mathcal{B}$ contains the PONs
that define the powers $c_{i}$ in Eq.~\eqref{Eq:F4}, i.e. 
\begin{eqnarray}
c_{1} & = & n_{\mu_{0}}(x_{0})-b_{0},\\
c_{2} & = & n_{\nu_{0}}(x_{0}+\hat{\mu}_{0})-b_{0},\\
c_{3} & = & n_{\mu_{0}}(x_{0}+\hat{\nu}_{0})-b_{0},\\
c_{4} & = & n_{\nu_{0}}(x_{0})-b_{0}.
\end{eqnarray}

Let us now discuss the integration of $F_4$ with the measure $\widetilde{\mathscr{D}U_{4}}$
defined in Eq.~(\ref{eq:haar41}). The integration over the links of the central
plaquette can be started by picking any of the links and for the function
$F_{4}$ one can reduce the integration to 
\begin{eqnarray}
I_{1}[g;b,c] & = & \int dU~\textrm{Tr}\left[Ug\right]^{b}\textrm{Tr}\left[U^{\dagger}\right]^{c}
\nonumber \\
 & = & \left[\partial_{x}^{b}\partial_{y}^{c} \int dU \, e^{x\textrm{Tr}\left[Ug\right]+y\textrm{Tr}\left[U^{\dagger}\right]}\right]_{\substack{x=0\\
y=0\\}},
\label{eq:Int_def}
\end{eqnarray}
where $U$ and $g$ are SU(2) matrices. Integrals of this type have
been computed in Ref.~\cite{Eriksson:1980rq}; they are given by 
\begin{eqnarray}
I_{1} & = & \left[\partial_{x}^{b}\partial_{y}^{c}\sum_{q=0}^{\infty}\frac{\left(xy\textrm{Tr}\left[g\right]+x^{2}+y^{2}\right)^{q}}{q!(q+1)!}\right]_{\substack{x=0\\
y=0\\}}.
\label{eq_I1}
\end{eqnarray}
For a non-vanishing result, the condition $2q=b+c$ must be fulfilled. The integral
$I_{1}$ is a polynomial in $\textrm{Tr}\left[g\right]$, i.e. 
\begin{eqnarray}
I_{1}[g;b,c] & = & \sum_{q=0}^{min\left(b,c\right)} \Gamma_{q}^{b,c} \, \textrm{Tr}\left[g\right]^{q},\label{eq:basic_int_sol}
\end{eqnarray}
where $\textrm{min}(b,c)$ stands for the minimum of $b$ and $c$, and
the coefficients $\Gamma_{q}^{b,c}$ are given by 
\begin{equation}
\Gamma_{q}^{b,c}=\delta_{\left\{ q\%2~b\%2\right\} }~\frac{b!~c!}{\left(\frac{b+c}{2}+1\right)!\,\left(\frac{b-q}{2}\right)!\,\left(\frac{c-q}{2}\right)!\,q!}~,\label{eq:coeff}
\end{equation}
and \%2 returns the remainder of the integer division by~2. 
The Kronecker delta in Eq.~(\ref{eq:coeff}) indicates that the polynomial in Eq.~(\ref{eq:basic_int_sol}) 
contains only odd or even powers of~$q$.
The evaluation of $I_{1}$ is a first step 
towards the evaluation of the weights $Q$. In our code the expression given in Eq.~(\ref{eq:basic_int_sol})
was used directly. The routine to compute $I_{1}$ was checked against a numerical evaluation of $I_{1}$ 
for a number of cases and both results agreed within machine precision. The integral $I_{1}$, given in
Eq.~\eqref{eq:basic_int_sol}, can be used recursively to perform the integration of Eq.~(\ref{Eq:F4}):
\begin{eqnarray}
\int\widetilde{\mathscr{D}U}_{4}\,F_{4} & = & \left(\frac{\beta}{N}\right)^{b_{0}}\frac{1}{b_{0}!}\sum_{q_{1}}^{\min\left(b_{0},c_{1}\right)}\Gamma_{q_{1}}^{b_{0},c_{1}}\sum_{q_{2}}^{\min\left(q_{1},c_{2}\right)}\Gamma_{q_{2}}^{q_{1},c_{2}}\nonumber \\
 & \times & \sum_{q_{3}}^{\min\left(q_{2},c_{3}\right)}\Gamma_{q_{3}}^{q_{2},c_{3}}\sum_{q_{4}}^{\min\left(q_{3},c_{4}\right)}\Gamma_{q_{4}}^{q_{3},c_{4}}.\label{eq:4link_func_sol}
\end{eqnarray}
Once the coefficients $\Gamma_{q}^{b,c}$ are known, one can get an approximate estimation
for the weights $Q$ and also for the transition probability $p$
which is defined in the Markov chain.

In principle, the calculation of the weights can be improved by considering more 
complex integrations over the gauge links as, for example: 
\begin{eqnarray}
I_{2} &=& \int dU \, \textrm{Tr}\left[Uv\right]^{a}\left[Ug\right]^{b}
\textrm{Tr}\left[U^{\dagger}\right]^{c} \nonumber \\
&=& \Biggl[ 
\partial_{x}^{a}\partial_{y}^{b}\partial_{z}^{c} 
\sum_{q=0}^{\infty} \frac{1}{q!(q+1)!} \bigl(
xy\textrm{Tr}\left[g^{\dagger}v\right] \nonumber \\
&& + \, xz \textrm{Tr}\left[v\right] + yz\textrm{Tr} [g]+x^{2}+y^{2}+z^{2}\bigr)^{q}
\Biggr]_{\substack{x=0\\y=0\\z=0}},
\label{eq:complicado}
\end{eqnarray}
However, the coding of this type of solutions in a Monte
Carlo simulation is rather complex and will not be pursued here. Alternatively
and keeping the same rationale as described so far, one can explore
integrations over more complex paths on the lattice.


\subsubsection{Integration over a long path
\label{sub:Improved-integration-update}}

The method described in the previous section can be extended to more
complex and longer lattice paths. From the practical point of view,
one has to compromise the length and complexity of the path to perform
the group integration with the coding of the outcome of group integration.

\begin{figure}[t]
\centering
\includegraphics[scale=0.4]{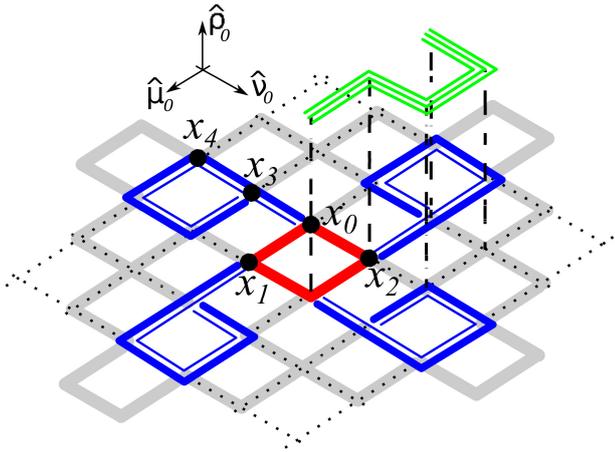} 

\caption{A 3-dimensional representation of the lattice. For an update of the
central plaquette $U_{\mu_0\nu_0}(x_0)$ (solid red lines), the links represented by red,
blue (double lines) and green (triple lines) are integrated in the computation of the
weight function ratio. In grey solid lines we shown a sample of the
links that are rotated to the identity in the calculation of $p$. See
text for details.}
\label{fig:3d} 
\end{figure}

Next, we consider an integration over the link variables which takes
into account a larger set of links that are decoupled from the remaining
lattice. In Fig.~\ref{fig:3d} we show, in color,
the links to be integrated in the computation of the probability transition  $p$. 
To avoid clutter, we do not
draw all links to be integrated exactly. In particular, the links
corresponding to the fourth dimension are not represented in the figure.
For the path represented in Fig.~\ref{fig:3d}, the links represented by solid red lines,
which belong to the central plaquette $U_{\mu_{0}\nu_{0}}(x_{0})$,
are integrated exactly together with those represented by double blue lines 
and by triple green lines.

The links in double blue lines are in the same plane as $U_{\mu_{0}\nu_{0}}(x_{0})$
and their integration involves terms which include the links of the
central plaquette. For example, the integrals referring to the links
belonging to the plaquettes $U_{\mu_{0}\nu_{0}}(x_{0})$ and $U_{\mu_{0}\nu_{0}}(x_{2})$
are not independent as these staples share $U_{\mu_{0}}(x_{2})$.
The same applies to the links belonging to the plaquettes $U_{\mu_{0}\nu_{0}}(x_{3})$
and $U_{\mu_{0}\nu_{0}}(x_{4})$, whose staples share $U_{\mu_{0}}(x_{3})$.
Also, the integration over the links in the $(\mu , \nu_{0})$ plane
are not independent of the integration of links in parallel planes
as those represented by triple green lines. In our integration over the longer path,
we will consider four green-type paths which belong to the upper parallel
plane shown in Fig~\ref{fig:3d}, on the down parallel plane and
similar paths related to the path which dislocated not by $\hat{\rho}_{0}$
but by the unit vector belonging to the fourth dimension not represented
in Fig~\ref{fig:3d}. The later three paths are not represented in
Fig~\ref{fig:3d}.

In the computation of the weights $Q$ and of the probability transition
$p$ the link variables represented in red (central plaquette with solid lines), blue (double lines) and green (triple lines) in Fig.~\ref{fig:3d}
are integrated exactly. Each of these link variables is coupled with
$2\left(d-1\right)$ staples which belong to $2\left(d-1\right)$
plaquettes. With the exception of the red, blue, and green links,
all staples associated with the links which are going to be integrated
are rotated to the identity matrix. In Fig.~\ref{fig:3d} the solid
lines in gray represent the link variables that are being fixed to
the identity matrix in the plane $\left(\mu_{0},\nu_{0}\right)$.
As in the integration described in Sec. \ref{sub:Simple-integration-update},
we are not building an exact maximal tree. Indeed, there are closed paths
whose links are all set to the identity. The approximation used to perform
the group integrations factorizes a local region, which is decoupled
from the remaining lattice, and, in this way, allows for an exact
group integration in each of the regions considered. Furthermore,
it factorizes the calculation of the weights $Q$, which enables an
easy estimation of the transition probability $p$.

Let us now discuss on how to integrate over the link variables in
color in Fig.~\ref{fig:3d}. In principle one can choose to start
the integration by considering any of the colored links. However,
we found that starting the integration by the links in green or the
links in blue and only then performing the integration of the links belonging
to the central plaquette $U_{\mu_{0}\nu_{0}}(x_{0})$ simplifies considerably
the integration process. For the integration over a path that is coupled
with the link $U_{\mu}(x)$ of the central plaquette, one can rely
on the result given in Eq. (\ref{eq:basic_int_sol}) applied recursively.
The outcome is a polynomial 
\begin{equation}
P[U_{\mu}(x)]=\sum_{k}\,\lambda_{k}\,\mbox{Tr}\left[U_{\mu}(x)\right]^{k},
\end{equation}
whose coefficients $\lambda_{k}$ are combinations of the coefficients
$\Gamma$ and are functions of the plaquette occupation numbers of
the region surrounding the integrated path. For the group integration,
every link belonging to the central plaquette is coupled with two
different paths, namely, the path in green which has four links and
the path in blue with five links. The total number of links to be
integrated is now forty and for this larger integration we define
$F_{40}\left[\mathcal{U},\mathcal{B},b_{0}\right]$ as being the local 
function of the approximate weight function ratio, see
Eq.~\eqref{eq:Q_ratio_approx}. The set of links $\mathcal{U}$ contains
all the forty gauge links to be integrated and the set of the plaquette
occupation numbers $\mathcal{B}$ include the $b_{\mu\nu}(x)$ whose
links are in the integrated paths. Recall that for the simpler
integration discussed previously a similar situation is found.

The formal expression for  $F_{40}\left[\mathcal{U},\mathcal{B},b_{0}\right]$
includes the central plaquette $U_{\mu_{0}\nu_{0}}(x_{0})$ and four
polynomials, one for each link variable $U_{l}\in U_{\mu_{0}\nu_{0}}(x_{0})$,
coming from the integration over the green and blue paths 
\begin{eqnarray}
F_{40} & = & \frac{1}{b_{0}!}\left(\frac{\beta}{N}\right)^{b_{0}}
\textrm{Tr}\left[U_{1}U_{2}U_{3}U_{4}\right]^{b_{0}}\prod_{l\in L}P_{\mathcal{B}\left(l\right)}[U_{l}],\label{eq:func_maxtree-1}
\end{eqnarray}
where $L$ is the set of coordinates of the links associated with $U_{\mu_{0}\nu_{0}}(x_{0})$,
see Eq.~\eqref{eq:coordenadas_plaquetta}. $P_{\mathcal{B}\left(l\right)}[U_{l}]$
is the polynomial coming from the integration of the green and blue
paths coupled to the link variable $U_{l}$, i.e. 
\begin{equation}
P_{\mathcal{B}\left(l\right)}[U_{l}]= P_{\mathcal{B}_{G}\left(l\right)}[U_{l}]\,P_{\mathcal{B}_{B}\left(l\right)}[U_{l}].\label{eq:polynomial_final}
\end{equation}
The polynomial $P_{\mathcal{B}_{G}\left(l\right)}$ is the outcome
of the integration over a green path and $P_{\mathcal{B}_{B}\left(l\right)}$
the outcome of integration over a blue path. The set $\mathcal{B}_{G}\left(l\right)$
includes the PONs of the plaquettes whose links
belong to the integrated green path. The set $\mathcal{B}_{B}\left(l\right)$
has the same meaning as $\mathcal{B}_{G}\left(l\right)$ but related
to a blue path. The union of $\mathcal{B}_{G}\left(l\right)$ and
$\mathcal{B}_{B}\left(l\right)$ defines the set $\mathcal{B}\left(l\right)$.
Finally, the set $\mathcal{B}$, required to perform
the group integration present in $F_{40}\left[\mathcal{U},\mathcal{B},b_{0}\right]$, is
given by the union of the four sets $\mathcal{B}\left(l\right)$ together
with the set of PONs of the plaquettes that share the links present
in $U_{\mu_{0}\nu_{0}}(x_{0})$.

Before providing expressions for $P_{\mathcal{B}_{G}\left(l\right)}$
and $P_{\mathcal{B}_{B}\left(l\right)}$ let us have a closer look
on the integrations leading to these polynomials.

\begin{figure}[t]
\centering
\includegraphics[scale=0.4]{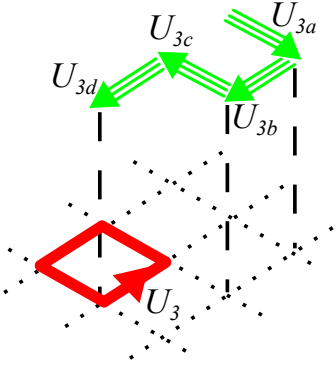}

\caption{Sublattice of the representation given in Fig.~\ref{fig:3d} with the  links labeled.}
\label{fig:3d_zoom-1} 
\end{figure}

In Fig.~\ref{fig:3d_zoom-1} the green path coupled to the link $U_{3}$
is shown in full detail. This path has four links $\left\{ U_{3a},U_{3b},U_{3c},U_{3d}\right\} $
and the integration over these links gives 
\begin{eqnarray}
P_{\mathcal{B}_{G}}[U_{3}] & = & \int\widetilde{\mathscr{D}U}_{G}\textrm{Tr}\left[U_{3}U_{3d}\right]^{b_{1}}
\textrm{Tr}\left[U_{3c}U_{3d}\right]^{b_{2}}\nonumber \\
& \times & \textrm{Tr}\left[U_{3a}U_{3b}U_{3c}\right]^{b_{3}}\mbox{Tr}\left[U_{3a}\right]^{c_{1}}\nonumber \\
& \times & \mbox{Tr}\left[U_{3b}\right]^{c_{2}}\textrm{Tr}\left[U_{3c}\right]^{c_{3}}
\textrm{Tr}\left[U_{3d}\right]^{c_{4}},
\label{eq:int_black}
\end{eqnarray}
and the integration measure reads 
\begin{equation}
\widetilde{\mathscr{D}U}_{G}= dU_{3a}\,dU_{3b}\,dU_{3c}\,dU_{3d}.\label{eq:haar_p1}
\end{equation}
The plaquette occupation numbers $\left\{ b_{1},b_{2},b_{3}\right\} $
refer to the plaquettes $\textrm{Tr}\left[U_{3}U_{3d}\right]$, $\textrm{Tr}\left[U_{3c}U_{3d}\right]$
and $\textrm{Tr}\left[U_{3a}U_{3b}U_{3c}\right]$, respectively, and
$\left\{ c_{1},c_{2},c_{3},c_{4}\right\} $, i.e. the powers of the
trace of the links that include $\left\{ U_{3a},U_{3b},U_{3c},U_{3d}\right\} $,
are given by sums of the plaquette occupation numbers similar to those
found in the case discussed in Sec.~\ref{sub:Simple-integration-update}.
For example, one has $c_{1}= n_{x_{0}+\hat{\nu}_{0}+\hat{\rho}-\hat{\mu}_{0},\nu_{0}}-\,b_{3}$.
The definition of the remaining $c_{i}$ and $b_{i}$ associated with
the integration over the green path is given in \ref{sec:Group-integration}.

The coefficients $\mathcal{B}_{G}=\left\{ b_{1},b_{2},b_{3},c_{1},c_{2},c_{3},c_{4}\right\} $
take into account the coupling of the central plaquette and a green
path attached to the link $U_{3}$. The degree and the coefficients
of the polynomial $P_{\mathcal{B}_{G}\left(l\right)}$ is determined
by the values of the $\mathcal{B}_{G}\left(l\right)$.

\begin{figure}[t]
\centering
\includegraphics[scale=0.8]{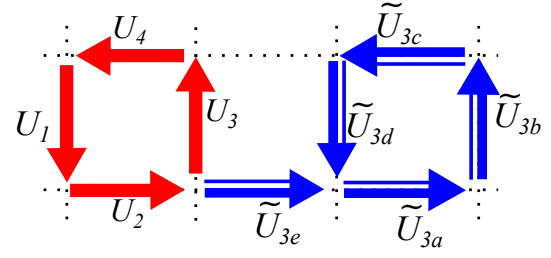}

\caption{Another sublattice of the representation given in Fig.~\ref{fig:3d} with the  links labeled.}
\label{fig:3d_zoom-2} 
\end{figure}

The blue path associated with the link $U_{3}$ is shown in Fig.~\ref{fig:3d_zoom-2}.
The blue path associated with the link $U_{3}$ includes the plaquette occupation numbers associated with the
first neighbor plaquette $\textrm{Tr}[U_{3}^{\dagger}\tilde{U}_{3e}\tilde{U}_{3d}^{\dagger}]$
of $U_{\mu_{0}\nu_{0}}(x_{0})$ and of its second neighbor $\textrm{Tr}[\tilde{U}_{3a}\tilde{U}_{3b}
\tilde{U}_{3c}\tilde{U}_{3d}]$.
The group integration over the blue path is 
\begin{eqnarray}
P_{\mathcal{B}_{B}}[U_{3}] & = & \int\widetilde{\mathscr{D}U}_{B}
\textrm{Tr}\left[\tilde{U}_{3a}\tilde{U}_{3b}\tilde{U}_{3c}\tilde{U}_{3d}\right]^{\tilde{b}_{2}}\nonumber \\
&& \times \, \textrm{Tr}\left[U_{3}^{\dagger}\tilde{U}_{3e}\tilde{U}_{3d}^{\dagger}\right]^{\tilde{b}_{1}}
\textrm{Tr}\left[\tilde{U}_{3a}\right]^{\tilde{c}_{1}}\textrm{Tr}\left[\tilde{U}_{3b}\right]^{\tilde{c}_{2}}
\nonumber \\
&& \times \, \textrm{Tr}\left[\tilde{U}_{3c}\right]^{\tilde{c}_{3}}
\textrm{Tr}\left[\tilde{U}_{3d}\right]^{\tilde{c}_{4}}\textrm{Tr}\left[\tilde{U}_{3e}\right]^{\tilde{c}_{5}} ,
\label{eq:int_blue}
\end{eqnarray}
for an integration measure given by 
\begin{equation}
\widetilde{\mathscr{D}U}_{B}=d\tilde{U}_{3a}d\tilde{U}_{3b}d\tilde{U}_{3c}\tilde{dU}_{3d}.
\label{eq:haar_p2}
\end{equation}
As for the green path, expressions for the coefficients
$\tilde{c}_{i},\tilde{b}_{i}$ are given in \ref{sec:Group-integration}.

The polynomials coming from performing the integrations over the green
and blue paths are computed in \ref{sec:Group-integration}.
It follows that for the green path
\begin{eqnarray}
P_{\mathcal{B}_{G}\left(l\right)}[U_{l}] & = & \sum_{q_{1}}^{\min\left(b_{3},c_{1}\right)}
\sum_{q_{2}}^{\min\left(q_{1},c_{2}\right)} 
\sum_{q_{3}}^{\min\left(b_{2},c_{3}+q_{2}\right)} \nonumber \\
&& \times \, \sum_{q_{4}}^{\min\left(b_{1},c_{4}+q_{3}\right)}
\Gamma_{q_{1}}^{b_{3},c_{1}}\,\Gamma_{q_{2}}^{q_{1},c_{2}}\,\Gamma_{q_{3}}^{b_{2},c_{3}+q_{2}}\nonumber \\
&& \times \, \Gamma_{q_{4}}^{b_{1},c_{4}+q_{3}} \textrm{Tr}\left[U_{l}\right]^{q_{4}},
\label{eq:int_green_sol}
\end{eqnarray}
while the blue path the group integration gives
\begin{eqnarray}
P_{\mathcal{B}_{B}\left(l\right)}[U_{l}] & = & \sum_{\tilde{q}_{1}}^{\min\left(\tilde{b}_{2},
\tilde{c}_{1}\right)} \, \sum_{\tilde{q}_{2}}^{\min\left(\tilde{q}_{1},\tilde{c}_{2}\right)}
\, \sum_{\tilde{q}_{3}}^{\min\left(\tilde{q}_{2},\tilde{c}_{3}\right)} \nonumber \\
&& \times \, \sum_{\tilde{q}_{4}}^{\min\left(\tilde{b}_{1},\tilde{c}_{4}+\tilde{q}_{3}\right)}\,
\sum_{\tilde{q}_{5}}^{\min\left(\tilde{q}_{4},\tilde{c}_{5}\right)} \, \Gamma_{\tilde{q}_{1}}^{\tilde{b}_{2},
\tilde{c}_{1}} \, \Gamma_{\tilde{q}_{2}}^{\tilde{q}_{1},\tilde{c}_{2}}\nonumber \\
&& \times \, \Gamma_{\tilde{q}_{2}}^{\tilde{q}_{2},
\tilde{c}_{3}}\,\Gamma_{\tilde{q}_{3}}^{\tilde{b}_{1},\tilde{c}_{4}+\tilde{q}_{3}}\,
\Gamma_{\tilde{q}_{5}}^{\tilde{q}_{4},\tilde{c}_{5}}\,\textrm{Tr}\left[U_{l}\right]^{\tilde{q}_{5}},
\label{eq:int_blue_sol}
\end{eqnarray}

In order to evaluate Eq.~\eqref{eq:polynomial_final} for each link of
the central plaquette $U_{l}\in\left\{ U_{1},U_{2},U_{3},U_{4}\right\} $,
it remains to multiply Eqs.~\eqref{eq:int_green_sol} and \eqref{eq:int_blue_sol}.
Once the polynomials $\,P_{\mathcal{B}\left(l\right)}[U_{l}]\,$ are
evaluated, we can integrate $F_{40}$, see Eq.~\eqref{eq:func_maxtree-1},
over the remaining links 
\begin{equation}
\int\widetilde{\mathscr{D}U_{4}}\,F_{40}=\int\widetilde{\mathscr{D}U_{4}}
\textrm{Tr}\left[U_{1}U_{2}U_{3}U_{4}\right]^{b_{0}}
\prod_{l\in L}P_{\mathcal{B}\left(l\right)}[U_{l}],
\label{eq:int_func2}
\end{equation}
and estimate the ratio between the weights $Q$ in order to evaluate
the probability transition $p$.

For the particular case of a local update transition $b_{0}\rightarrow b_{0}\pm\Delta$,
the polynomials $P_{\mathcal{B}\left(l\right)}[U_{l}]$ contributing
to $F_{40}$ do not depend on $\Delta$, i.e. on the update of the
central plaquette and, therefore, they do not need to be evaluated
twice to compute $p$. Note that the function $F_{40}$ defined in Eq.~\eqref{eq:func_maxtree-1}
is given by a sum of terms like $F_{4}$ given in Eq.~\eqref{Eq:F4}.
It follows that the solution of the group integration in Eq.~\eqref{eq:int_func2}
is a sum of the solutions that look like Eq.~\eqref{eq:4link_func_sol}.
Then, the group integration is reduced to the computation of factorial
numbers and, it follows from the definition and the approximation
used, that the transition probability is a real and positive definite
number.

\subsection{Nonlocal update\label{sub:Nonlocal-update}}

The Monte Carlo updates discussed in Secs. \ref{Sec:groupinte}, 
\ref{sub:Simple-integration-update} and \ref{sub:Improved-integration-update} do not allow us to access
all possible configurations for the plaquette occupation numbers.
For example, those local updates are unable to change a given plaquette
occupation number from an odd natural number to an even natural number
or vice-versa. As discussed in Sec.~\ref{sub:Constraint-over-the}, the
introduction of a global or a nonlocal update can improve the algorithm
in the sense that it enlarges the space sampled by the algorithm.

A nonlocal update can be implemented via a simultaneous transformation
of all the plaquette occupation numbers over a plane surface, where
each of the PONs is changed accordingly to $b_{\mu\nu}(x)\rightarrow b_{\mu\nu}(x)\pm\Delta$,
where $\Delta$ is not necessarily a multiple of N. For this update,
the number of links to be integrated increases with the lattice size.
Recall that for the updates discussed previously, the number of links
integrated to compute the weights depends only on the type of update
and is fixed \textit{a priori} for each of the updates. Although by enlarging
the size of the space sampled by the algorithm, this nonlocal update might
not be enough to guarantee full ergodicity of the algorithm, but it
certainly helps in approaching an ergodic update.
Of course, one can introduce other types of nonlocal updates as, for example, 
an update of the PONs attached to a cube.
The updating process where a plane surface is filled with PONs that are not 
multiples of N can not generate a configuration where the PONs that are not 
multiples of N are attached to the cube surface.
In addition to the so-called planar update, and to comply with full ergodicity,
one should also implement the cube type of update. However, its
implementation is rather complex and its impact on the performance of the
algorithm will be the object of a future report.

\begin{figure}[t]
\centering
\includegraphics[scale=0.7]{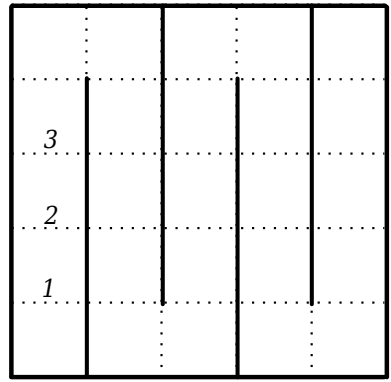}
\caption{Two-dimensional representation of the $5\times5$ lattice with periodic
boundary conditions. The solid lines are fixed to the identity and the
doted ones are integrated in the evaluation of the weight function
ratio corresponding to a nonlocal plane update.}
\label{fig:plane_gauge_fix} 
\end{figure}

Let us now discuss the group integration to compute the transition
probability $p$. In Fig.~\ref{fig:plane_gauge_fix} we show the
surface over which the plaquette occupation numbers are to be updated.
In order to perform the group integration, the links represented by
solid lines are set to the identity matrix and those represented by
doted lines are to be integrated exactly for the weight evaluation.
In $d>2$ dimensions and in what respects the group integration, the
links in Fig.~\ref{fig:plane_gauge_fix} are coupled with staples
in perpendicular planes. In the integration to compute the transition probability
for this nonlocal update, all those staples are set to the identity
matrix. Again, we are not building an exact  maximal tree but the approximation
allows us to get  relatively simple expressions in the calculus of the transition 
probability $p$.

The local function $F_{p}$, associated with the update over a $5\times5$
plane represented in Fig.~\ref{fig:plane_gauge_fix}, contains 24
link variables and is given by 
\begin{eqnarray}
F_{p} & = & \frac{\text{Tr}\left[U_{1}\right]^{b_{1}}}{b_{1}!}~\frac{\text{Tr}
\left[U_{1}^{\dagger}U_{2}\right]^{b_{2}}}{b_{2}!}~\frac{\text{Tr}
\left[U_{2}^{\dagger}U_{3}\right]^{b_{3}}}{b_{3}!}\nonumber \\
& & \times \, \cdots \, \frac{\text{Tr}\left[U_{23}^{\dagger}U_{24}\right]^{b_{24}}}{b_{24}!}
\frac{\text{Tr}\left[U_{24}\right]^{b_{25}}}{b_{25}!}~\text{Tr}\left[U_{1}\right]^{c_{1}}\nonumber \\
&& \times \, \text{Tr}\left[U_{2}\right]^{c_{2}}~\text{Tr}\left[U_{3}\right]^{c_{3}}~\dots~\text{Tr}\left[U_{24}\right]^{c_{24}},
\end{eqnarray}
where all $b_{i}$ are plaquette occupation numbers belonging to the
plane where the nonlocal update takes place, the $c_{i}$ are related
to the integrated link $U_{i}$ and are given by a sum of plaquette
occupation numbers belonging to the plaquettes that share $U_{i}$
in other planes than the updated plane.

Starting the group integration by the link labelled $1$ in Fig.~\ref{fig:plane_gauge_fix},
the integration function is of the same type as that defined in Eq.~\eqref{eq:Int_def}
and whose solution is given in Eq.~\eqref{eq:basic_int_sol}. The integration
leads to a polynomial of the trace of the link with label $2$. For
the integration of the link labelled $2$ one uses the solution in Eq.~\eqref{eq:basic_int_sol}
and repeat the process to the subsequent links, as for the integration
of the green and blue paths in the local update associated with Fig.~\ref{fig:3d}.

\section{Observables\label{sec:Observables}}

We implemented the algorithm to compute the mean value of the plaquette
and the mass of the $J^{PC}=0^{++}$ glueball. The mean value of the plaquette is easily 
computed in terms of the plaquette occupation numbers. In the partition function given
by Eq.~\eqref{eq:partition_func}, the plaquette $U_{\mu\nu}(x)$ comes associated with the 
factor $\beta$. Formally, one can identify a different $\beta$ with each of the plaquettes
and make the replacement $\beta\rightarrow\beta_{\mu\nu}(x)$. Ignoring
the constant $C$ in Eq.~\eqref{eq:partition_func}, it follows that
\begin{eqnarray}
\frac{\partial\textrm{ ln }Z}{\partial\beta_{\mu\nu}(x)} 
&=& \frac{1}{Z}\int\mathscr{D}U\left(\frac{\textrm{Re Tr}\left[U_{\mu\nu}(x)\right]}{N}\right)W\left[U\right], \nonumber \\
& = & \left\langle N^{-1}\textrm{Re Tr }U_{\mu\nu}(x)\right\rangle,
\label{eq:partition_func_derivative} 
\end{eqnarray}
where $W\left[U\right]$ is the Boltzmann weight factor in the standard
representation of the partition function. Performing the same operations
with the partition function written in the new representation, 
given by Eqs.~\eqref{eq:partition_Function_New} and \eqref{eq:weight_Q}, it follows 
that 
\begin{equation}
\frac{\partial\textrm{ ln }Z}{\partial\beta_{\mu\nu}(x)}  
= \frac{1}{Z}\sum_{\left\{ b\right\} }\left(\frac{b_{\mu\nu}(x)}{\beta}\right)Q_{\left\{ b\right\} },
\end{equation}
and, therefore, the plaquette expectation value is given by the average
of the dual variables $b_{\mu\nu}(x)$ 
\begin{equation}
\left\langle \frac{\textrm{Re Tr }U_{\mu\nu}(x)}{N}\right\rangle 
= \left\langle \frac{b_{\mu\nu}(x)}{\beta}\right\rangle .\label{eq:plaq_expect}
\end{equation}
The average of the plaquette expectation value over the lattice volume,
called plaquette mean value $u$, reads 
\begin{equation}
u=\frac{1}{V_{p}}\left\langle \frac{\sum\limits _{x,\mu\nu}b_{\mu\nu}(x)}{\beta}\right\rangle .
\label{eq:plaq_mean}
\end{equation}
The value of $u$ estimated with our algorithm will be compared
with the output of a conventional heat bath Monte Carlo method.

In this exploratory work besides the mean value of the plaquette we
also compute the mass of the scalar glueball with quantum numbers
$J^{PC}=0^{++}$. This requires the building of an interpolating field
$\Phi$ with the right quantum numbers and writing $\Phi$ in terms
of the dual variables $b_{\mu\nu}(x)$. A first step toward
the computation of the mass of the scalar glueball is the evaluation of the 
correlation function 
\begin{eqnarray}
G(x-y)=\Bra{0}\Phi(x)\Phi(y)\Ket{0}.
\label{eq_propagator}
\end{eqnarray}
Setting $y$ fixed at the origin of the lattice, then the zero momentum
Euclidean space Green's function reads
\begin{eqnarray}
G(t) & = & \frac{1}{4\pi^{2}}\int_{0}^{\infty}dp~\frac{p^{2}}{\sqrt{p^{2}+m^{2}}}e^{-\sqrt{p^{2}+m^{2}}t}
\nonumber \\
& = & \frac{m^{2}}{4\pi^{2}}\int_{1}^{\infty}dz~\sqrt{z^{2}-1}e^{-zmt},
\end{eqnarray}
where $m$ is the mass of the glueball ground state and the last line
is the result of making the change of variable $p^{2}+m^{2}=z^{2}m^{2}$
in the first line. The integration over $z$ is given in terms of 
the $K_{1}$ Bessel:
\begin{eqnarray}
G(t)  =\frac{(1/2)!}{2\pi^{5/2}}\frac{m}{t}K_{1}(mt)
\approx\frac{\sqrt{m}}{t^{3/2}}e^{-mt}\bigg(1+\mathcal{O}(1/t)\bigg),
\label{eq:correlation}
\end{eqnarray}
where the second expression holds for large values of $t$.

The lattice version of the operator $\Phi$ is constructed by mapping
the continuum symmetries and therefore the quantum numbers of the
corresponding particle into the hypercubic group \cite{Berg:1982kp}.
For the ground state and for the channel $J^{PC}=0^{++}$, the simplest
operator is given by 
\begin{eqnarray}
\Phi(t) & = & \sum_{\vec{x}}\sum_{\substack{\mu<\nu\\
\nu\neq t
}
}\frac{\textrm{{\rm Re} {\rm Tr }}U_{\mu\nu}(\vec{x},t)}{N},
\label{eq:glue_op}
\end{eqnarray}
i.e., the sum of spacelike plaquettes. With the use of Eq.~\eqref{eq:plaq_expect},
the operator $\Phi$ can be mapped into the new representation
and is given in terms of $b_{\mu\nu}(x)$ as
\begin{eqnarray}
\Phi(t) & = & \sum_{\vec{x}}\sum_{\substack{\mu<\nu\\
\nu\neq t
}
}\frac{b_{\mu\nu}(\vec{x},t)}{\beta}.\label{eq:glue_op-1}
\end{eqnarray}

The estimation of the glueball masses from correlation functions of
type given in Eq.~\eqref{eq:correlation} with smaller statistical errors
is not an easy task. Indeed, given that $G(t)$ decays exponentially
with Euclidean time, the signal to noise ratio decreases speedily
for large Euclidean time and, therefore, on the lattice one can only
rely on a limited number of time slices to estimate $m$. Although
there are a number of techniques to improve the signal to noise ratio,
as e.g. the use of anisotropic lattices or the use of smeared operators
\cite{Chen:2005mg,Teper:1987wt,Albanese:1987ds}, we will take
the interpolating operator as given in Eq.~\eqref{eq:glue_op}, with
the representation given in Eq.~\eqref{eq:glue_op-1}, to test the
algorithm.

In practice, for estimating the scalar glueball mass, a number of
uncorrelated configurations will be generated and the operator $\Phi$
will be computed using Eq.~\eqref{eq:glue_op-1}. From the interpolating
field we evaluate the scalar glueball connected Green function 
\begin{equation}
G(t)= \frac{1}{T}\sum_{\tau} \left[\Braket{\Phi(t)\Phi(t+\tau)}-\Braket{\Phi(t)}\Braket{\Phi(t+\tau)}
\right],\label{eq:correlation_lat}
\end{equation}
where $T$ is the lattice time length and the second term on the right
hand side in Eq.~\eqref{eq:correlation_lat} removes the vacuum contribution
to the signal. The mass of the $J^{PC}=0^{++}$ glueball 
is measured fitting the lattice estimation in Eq.~\eqref{eq:correlation_lat}
to the functional form given in Eq.~\eqref{eq:correlation}.


\section{Results\label{sec:Monte-Carlo-results}}

In the simulations we start the Markov chain
with a cold start, where all $b_{\mu\nu}(x)=0$, and the Monte Carlo
updates use both the local and nonlocal updates.

For the local updates, a given PON $b_{\mu\nu}(x)$ is chosen randomly and a 
change by $\pm2$ is proposed with the sign being chosen randomly. This process 
is repeated $V_{p}$ times, where $V_{p}$ is the total number of lattice plaquettes. 
To this set of updates we call one Monte Carlo step or full sweep for the local update.

For the nonlocal update, a two dimensional surface is chosen randomly
on the lattice and for each PON $b_{\mu\nu}(x)$ on the surface a change 
by $\pm1$ is proposed randomly. The process is repeated $N_{p}$ times, where $N_{p}$ is 
the number of two dimensional surfaces on the lattice. To this set of updates we call 
one Monte Carlo step or full sweep for the nonlocal surface update.


\subsection{Sampling and the mean value of the plaquette}

\begin{figure}[t]
\centering
\includegraphics[scale=0.4]{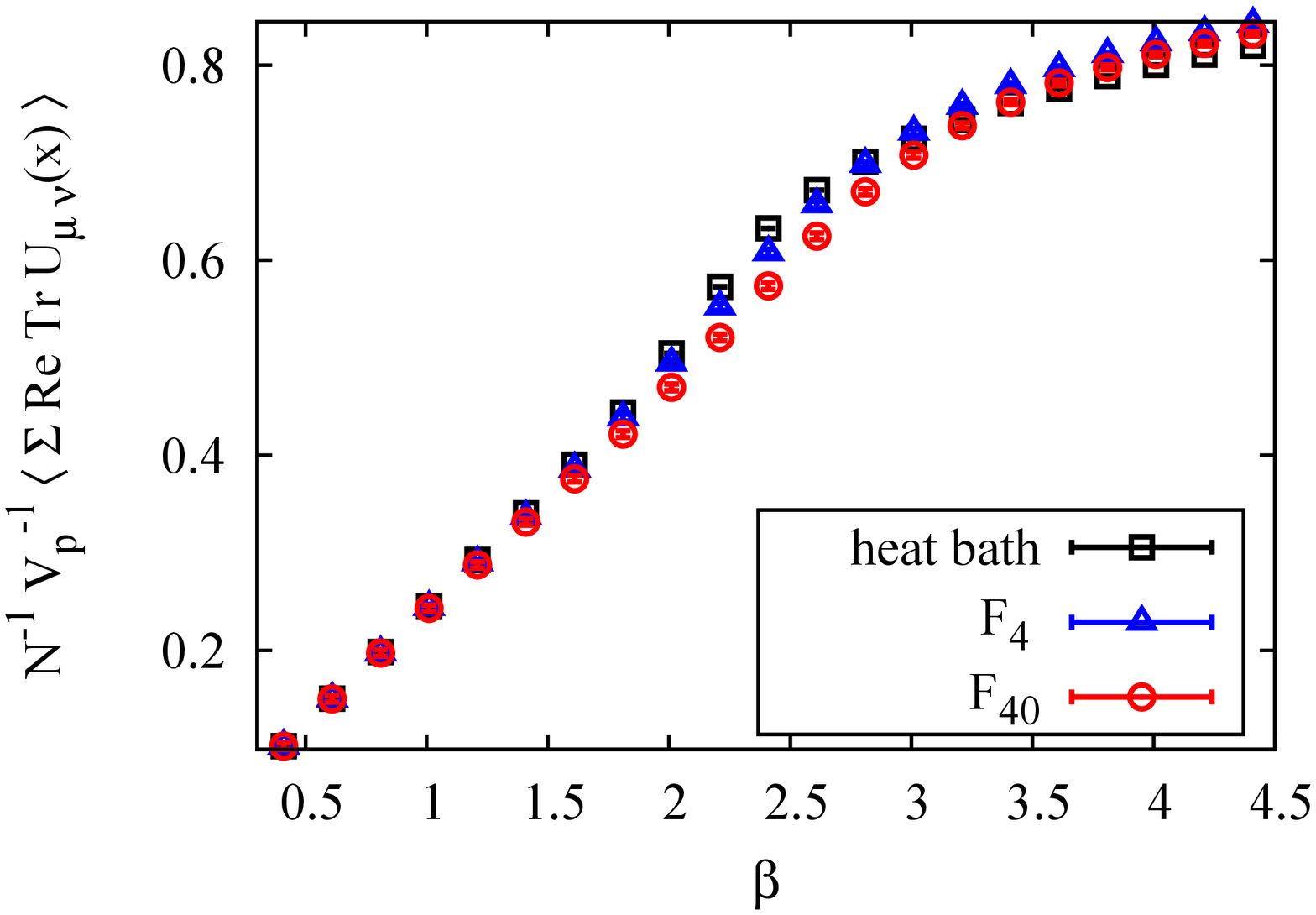} \\[0.5true cm]
\includegraphics[scale=0.4]{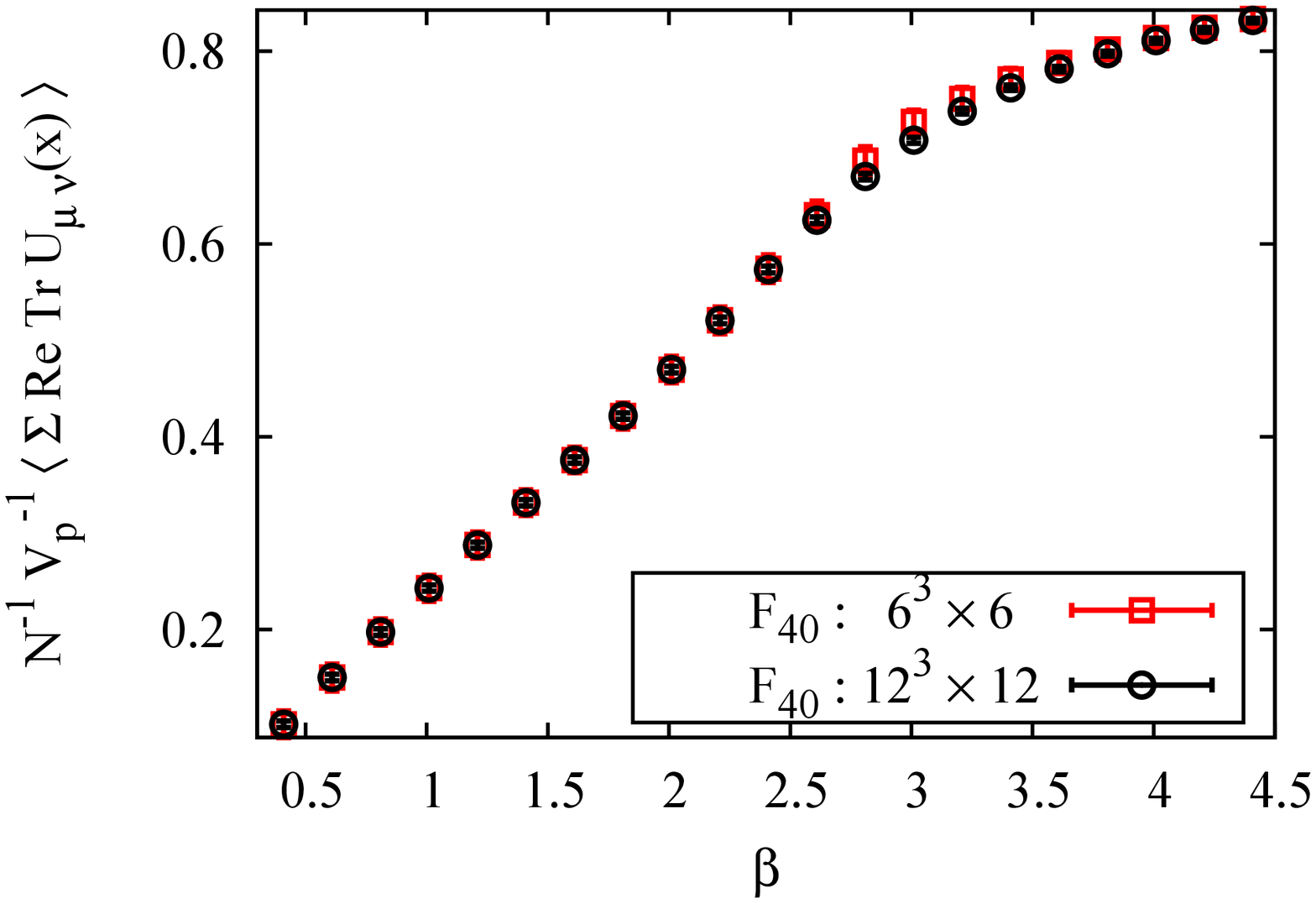}

\caption{Mean value of the plaquette from simulations using the algorithm
and the heat bath algorithm. Top panel: Comparison of the 
performance of the two local updates. Lower panel: Finite volume effects when using 
the the present algorithm.}
\label{fig:plaq_mean} 
\end{figure}

For the evaluation of the mean value of the plaquette given in terms of the dual 
variables, as given in Eq.~\eqref{eq:plaq_mean}, we simulate two
different lattice volumes, $6^{4}$ and $12^{4}$, for various values of $\beta$.
For each of the simulations, after discarding $10^{3}$ combined Monte
Carlo steps for thermalization, we consider $10^{4}$ configurations
separated by $10$ combined Monte Carlo steps. Our numerical experiments
have shown that a separation of 10 combined Monte Carlo sweeps is
enough to decorrelate the observables measured in the current work.

In Fig.~\ref{fig:plaq_mean} we compare the results obtained with the present algorithm 
with the results obtained with the heat bath algorithm (also for the standard Wilson
action) implemented with the library Chroma~\cite{Edwards:2004sx}.  The results shown 
for the heat bath algorithm refer to simulations performed on a $10^{4}$ lattice, for an 
ensemble with $10^{4}$ configurations, separated by $5$ Monte Carlo steps.

\begin{figure}
\centering
\includegraphics[scale=0.33]{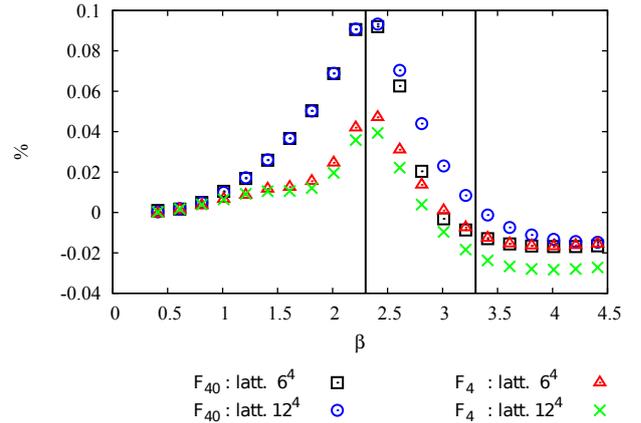} 
\par
\caption{The same as in Fig.~\ref{fig:plaq_mean} but for the relative deviation
of the results obtained with present algorithm with respect to the heat bath results. 
The vertical black lines show the interval of $\beta$ values where the efficiency
of the nonlocal update is higher. See text for details.}
\label{fig:comparativo} 
\end{figure}

In the top panel of Fig.~\ref{fig:plaq_mean} we show the plaquette mean value
obtained in simulations of a $12^4$ lattice combining the local and nonlocal updates 
against the results of the standard Wilson action using a heat bath simulation. 
As can be seen, there is good agreement between the results obtained with our algorithm, 
using any of the local updates, and with those obtained with the heat bath simulation 
in the strong coupling limit. The data also suggest that in the weak coupling limit the 
present algorithm prediction for $u$ converges to the value given by the heat bath
algorithm.

In what concerns the $\beta$ dependence of the results, the present prediction 
for $u$ starts to deviate from the heat bath result for $\beta\sim1.5$ up to $\beta\sim3$,
but its maximal deviation is about 0.1\%  and occurs for $\beta\sim2.3$. Interestingly, 
in this range the local algorithm which takes into account the small\-er number of integrations, 
see Sec.~\ref{sub:Simple-integration-update}, is closer to the results of the 
heat bath simulation. However, as one approaches the continuum limit, i.e. for $\beta\gtrsim3$, 
it is the algorithm which uses the other local update, 
see Sec.~\ref{sub:Improved-integration-update}, which is closer to the heat bath outcome. 
Indeed, for the algorithm whose local update takes into account the larger number 
of group integrations the deviations from the heat bath result are marginal
for $\beta\gtrsim3$.

The volume dependence of the algorithm can be seen in the lower panel of
Fig.~\ref{fig:plaq_mean}, where the sampling of $u$ is investigated for two 
different lattice volumes. Recall that the full Monte Carlo update is defined 
by a combination of local and nonlocal updates. The data show no or only 
a mild dependence on the lattice volume. 

In Fig.~\ref{fig:comparativo} we show the relative
deviation of the present estimation of $u$ with respect to
the heat bath results for different lattice volumes. The deviations
are negligible in the strong coupling limit and are very small when the continuum 
limit is approached. The maximal deviations $\lesssim 0.1$\% occur for intermediate 
values of the coupling around $\beta \sim 2.3$.
From the  figure one can also read the  improvement of considering 
$F_{40}$ instead of $F_{4}$; recall that $F_{40}$ takes into account forty Haar 
integrals in the evaluation of $p$ while $F_{4}$ takes only four Haar integrals. 
In particular, for the largest lattice, the data show a notorious improvement on the values 
of the mean value of the plaquette relative to the heat bath numbers when using
$F_{40}$. Indeed, in the continuum limit, when one uses $F_{40}$ to estimate $p$, the 
deviations are about $\sim 50\%$ smaller compared to computations using $F_{4}$.

The numerical simulations performed show that our approach is a good approximation in the strong 
and weak coupling limits.
Given that Boltzmann weights are proportional to powers $\beta$, in the strong coupling limit, 
i.e. for small $\beta$ values, most likely the dual variables are zero or close to zero and setting 
the link variables to the identity matrix is essentially an irrelevant operation.
As $\beta$ increases the dual variables start to deviate from zero, the previous argument no 
longer applies, and one can expect deviations from the exact result. This seems to be the case
for $\beta$ values in the range $2.5 < \beta < 3$. Although one would expect that integrating 
more links is always better, one should recall that the integration is not exact. While it is 
true that the $F_{40}$ updates take into account more link variables, it also sets a large number 
of link variables to the identity matrix and, therefore, at some stage it can become less 
accurate than the $F_4$ update. On the other hand, as the continuum limit is approached, the link 
variables approach the identity and, in this case, our approximation reproduces faithfully 
the theoretical expectations.

\begin{figure}[t]
\centering
\includegraphics[scale=0.4]{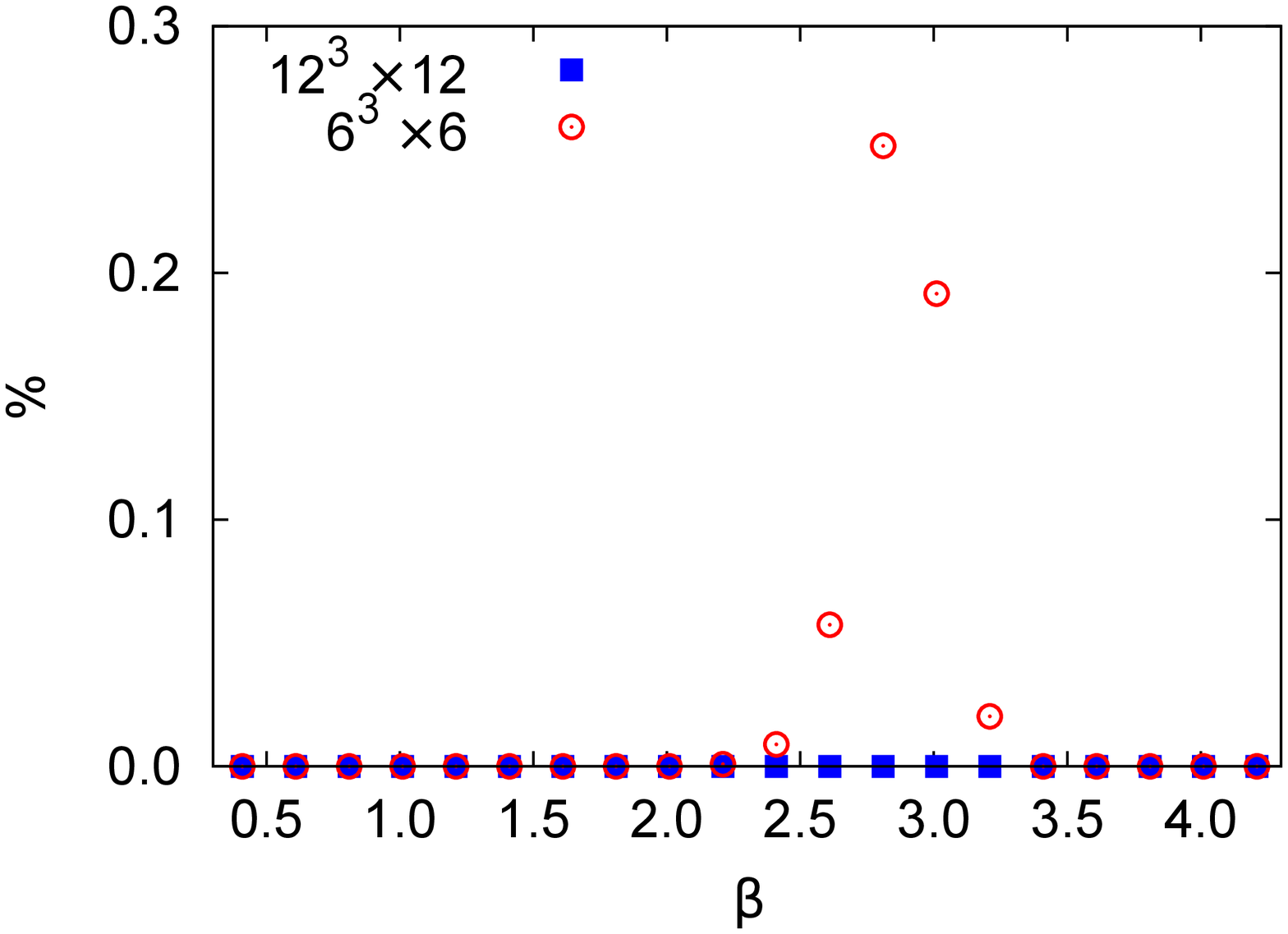} \\
\includegraphics[scale=0.4]{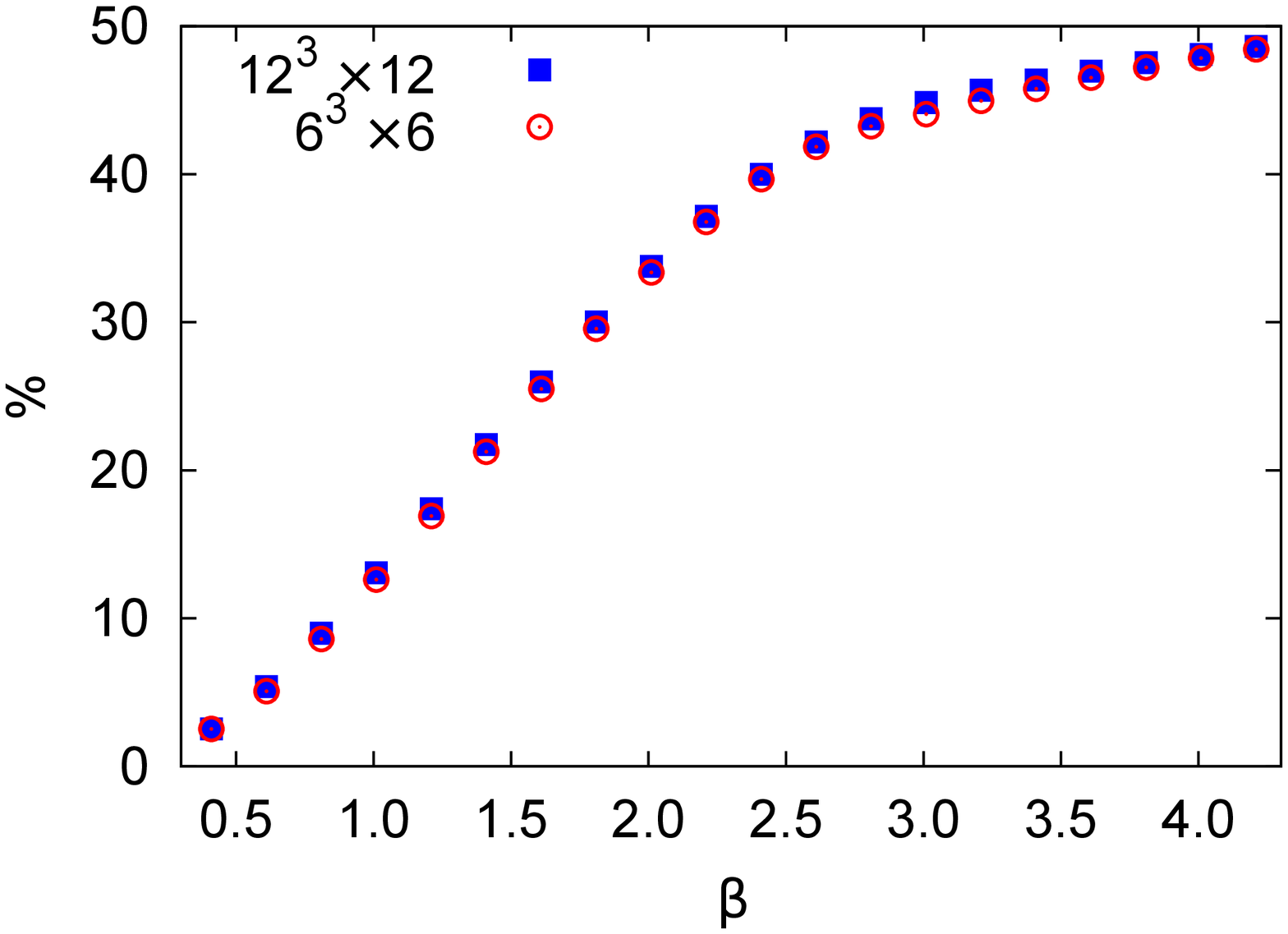} 

\caption{Efficiency of the algorithm for the various updates. Top panel: Nonlocal update, 
see Sec.~\ref{sub:Nonlocal-update}. Bottom panel: Local updates using the function $F_{40}$, 
see  Sec.~\ref{sub:Improved-integration-update}}
\label{acceptances} 
\end{figure}

The performance of our algorithm for different $\beta$ values
and different volumes can be understood looking at the update efficiency
$\mathscr{E}$, defined as its acceptance rate in the Markov chain.
As can be seen in the top panel of Fig.~\ref{acceptances}, the nonlocal update
has an essentially vanishing $\mathscr{E}$, with the exception of
the smaller lattice and over a narrow range of $\beta$ values. Note,
however, that the region where the simulations using the two volumes
give different values for $u$, see the lower panel in Fig.~\ref{fig:plaq_mean},
is precisely the region of $\beta$ where the efficiency associated
with the nonlocal update has its maximum value. Furthermore, the results
of Figs.~\ref{fig:plaq_mean} and~\ref{acceptances} suggest that
the nonlocal update plays an important role. Indeed,
for the smaller lattice volume and for $\beta$ in the range $2.5-3$,
the efficiency $\mathscr{E}$ is maximal and non negligible for the
nonlocal update, which makes the estimation of $u$ by the present  algorithm
closer to the values provided by the heat bath method.
The deviations of our estimation for $u$ relative to the heat bath result
for the smaller volumes are milder for $\beta$ in the range $2.5-3$, as can
be seen from Fig.~\ref{fig:comparativo}. For the larger volumes, $\mathscr{E}$ is
always residual and the our estimation of $u$ shows larger
deviations which are, nonetheless, less than 0.1\% relative to the heat 
bath numbers.

Herein, we considered a single type of nonlocal update but many other
possibilities can be explored to achieve a better and more complete
sampling of the dynamical range of values allowed for the plaquette
occupation number space. Within the rationale considered in this work,
the building of a algorithm, i.e. the implementation of other
types of nonlocal updates, implies a compromise between a given geometrical
setting, i.e. the definition of a given set of links over a large
region of the lattice, and the ability of being able to perform the
group integration over the corresponding sublattices. Recall that,
within our framework, the local
updates do not sample the entire $\{b\}$ space. For example, for
the local updates the $\{b\}$ remain either in the subset of odd
or even natural numbers. The nonlocal updates were built to allow
for a better dynamical range, allowing for transitions in Markov chain
where the PONs could become either odd or even natural numbers.
The search for other nonlocal types of updates is one of the features
that we aim to explore in a future work.

Another way of reading the results in Fig.~\ref{acceptances} is that one should 
improve the efficiency $\mathscr{E}$ of the non local update. Indeed, for the 
local updates its acceptance rate is always $\sim10\%$ or above, reaching a 
value of about $50\%$ as the continuum limit is approached. On the other hand, 
the nonlocal update defined in Sec.~\ref{sub:Nonlocal-update}, has an 
extremely low $\mathscr{E}$, with has a maximum of $\sim0.25\%$ for 
$\beta\sim2.7$ for the smaller lattice and being always residual for the 
larger lattice. The low values 
for the efficiency associated with the nonlocal update mean that the 
plaquette occupation numbers are essentially trapped into the subset of the
odd or the even natural numbers which is sampled by the local
Monte Carlo updates.

The local updates change, in a single update, a fixed number
of $b_{\mu\nu}(x)$ and, in principle, are not so sensitive to volume
effects as the nonlocal updates which are affected by surface
effects.


\subsection{$J^{PC}=0^{++}$ glueball mass}

In order to estimate the $J^{PC}=0^{++}$ glueball mass
we simulate the theory for $\beta=3.01$ on a $10^{3}\times20$
lattice, with a Monte Carlo step combining the local update
as defined in Sec.~\ref{sub:Improved-integration-update} and 
nonlocal updates as defined in Sec.~\ref{sub:Nonlocal-update}.

For the conversion of the glueball mass into physical units, we rely on Ref.~\cite{Bloch:2003sk}
which uses the string tension $\sqrt{\sigma}=440$ MeV and assumes
\begin{equation}
\textrm{ln}\left(\sigma a^{2}\right) = -4\frac{4\pi^{2}}{\beta_{0}}\beta
+ \frac{2\beta_{1}}{\beta_{0}^{2}}\textrm{ln}\left(\frac{4\pi^{2}}{\beta_{0}}\beta\right)
+ \frac{4\pi^{2}}{\beta_{0}}\frac{d}{\beta}+c,
\end{equation}
where the first two terms are the predictions of 2-loop perturbation
theory and the remaining terms parameterize higher-order effects.
The parameters $c=4.38(9)$ and $d=1.66(4)$ were set by fitting the
lattice data for the string tension using simulations with $\beta\in[2.3,2.85]$.
For $\beta=3.01$, the above relation estimates $a\approx0.02$ fm
for the lattice spacing.

\begin{figure}[t]
\centering
\includegraphics[scale=0.33,angle=0]{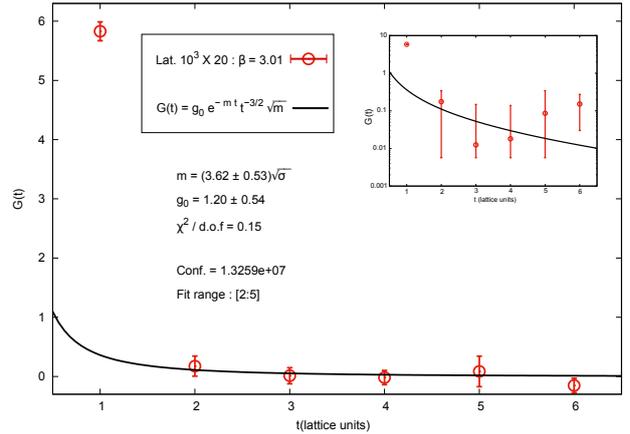} 

\caption{Lattice estimation of correlation function $G(t)$. The curve in
black is the fit of the lattice data to the asymptotic expression in 
Eq.~\eqref{eq:fit_func} and for the fitting range $t\in[2,5]$. In the inset,
$G(t)$ in logarithm scale.}
\label{glueball_fit} 
\end{figure}

The glueball mass is evaluated from the asymptotic expression
for the two-point correlation function 
\begin{equation}
G(t)~=~g_{0}\,\frac{\sqrt{m}}{t^{3/2}}\,e^{-mt}\ .\label{eq:fit_func}
\end{equation}
Our lattice estimations for $G(t)$ use $\sim10^{7}$ configurations
and the correlation function can be seen in Fig.~\ref{glueball_fit}.
Despite using a large ensemble, our Monte Carlo code is not parallelized.
However, the ensembles were built running the code on various independent
standalone machines. For $t\geqslant6$ the lattice two point correlation
function becomes negative and compatible with zero within one standard
deviation and, therefore, lattice Euclidean times larger than~6
will not be considered.

\begin{table*}
\caption{Fits of the glueball correlation function with the reduced chi squared
being $\nu^{2}=\chi^{2}/d.o.f.$}
\begin{tabular*}{\textwidth}{@{\extracolsep{\fill}}ccc@{}}
\hline 
\begin{tabular}{c}
\tabularnewline
Range\tabularnewline
\end{tabular} & %
\begin{tabular}{c}
Conf. = $1.12357\times10^{7}$\tabularnewline
\begin{tabular}{ccc}
$m/\sqrt{\sigma}$\quad\quad & $g_{0}$ & \quad\quad$\nu^{2}$\tabularnewline
\end{tabular}\tabularnewline
\end{tabular} & %
\begin{tabular}{c}
Conf. = $1.3259\times10^{7}$\tabularnewline
\begin{tabular}{ccc}
$m/\sqrt{\sigma}$\quad\quad & $g_{0}$ & \quad\quad$\nu^{2}$\tabularnewline
\end{tabular}\tabularnewline
\end{tabular}\tabularnewline
\hline 
\begin{tabular}{c}
{[}2:4{]}\tabularnewline
{[}2:5{]}\tabularnewline
{[}2:6{]}\tabularnewline
\end{tabular} & %
\begin{tabular}{ccc}
3.62(98) & 1.11(54) & 0.13\tabularnewline
3.61(86) & 1.15(48) & 0.10\tabularnewline
3.61(00) & 0.97(91) & 0.36\tabularnewline
\end{tabular} & %
\begin{tabular}{ccc}
3.62(45) & 1.17(61) & 0.19\tabularnewline
3.62(53) & 1.20(54) & 0.15\tabularnewline
3.61(20) & 0.96(99) & 0.56\tabularnewline
\end{tabular}\tabularnewline
\hline 
\end{tabular*}
\label{tab:fit_glueball_mass}
\end{table*}


The correlation function for $t=1$ does not comply
with the remaining values for larger $t$ and with Eq.~\eqref{eq:fit_func}
and, therefore, in the estimation of $m$ it is discarded. In
the measurement of the glueball mass we consider three different fitting
ranges and two large and independent ensembles as described in Tab.~\ref{tab:fit_glueball_mass}.
For each of the fitting ranges considered, the lattice data are well
described by the asymptotic expression for the correlation function in Eq.~\eqref{eq:fit_func},
as can be seen by the values of the $\chi^{2}/.d.o.f.$. Furthermore,
$m$ and $g_{0}$ are independent of the fitting range. The simulations
point towards a glueball mass of $1588\pm378$ MeV for a $\sqrt{\sigma}=440$
MeV.

The simulation described so far uses a small physical volume and the
simplest operator to estimate the glueball mass.
Indeed, none of the available techniques to improve the signal to
noise ratio is used in the numerical experiments. However, despite
this limitations we are able to reproduce the numbers that can be
found in the literature.
Our estimates for the intermediate fitting 
range are $m=(3.61\pm0.86)\,\sqrt{\sigma}$ and $m=(3.62\pm0.53)\,\sqrt{\sigma}$, and 
agree within one standard deviation with the numbers quoted above.
In Ref.~\cite{Falcioni:1982ja}
the authors report several estimates for SU(2) scalar glueball
mass.
Ref.~\cite{Berg:1980gz} uses the same interpolating
operator for the glueball as ours and reports the value $m=(3.7\pm1.2)\,\sqrt{\sigma}$, 
as we can see, the error in our estimation is about 30\% smaller.
In Ref.~\cite{Teper:1987wt}, the simulation is done using improved signal
to noise methods and the authors report the value $m=(3.12\pm0.22)\,\sqrt{\sigma}$.
Finally, the more recent calculation in Ref.~\cite{Lucini:2004my}
gives $m=(3.78\pm0.07)\,\sqrt{\sigma}$. 

\section{\label{sec:Conclusions}Summary}

In the present work we discuss a mapping of the lattice Wilson action into an approximate dual 
representation, whose dynamical variables, the so-called  plaquette occupation numbers (PONs)
$\left\{ b_{\mu\nu}\left(x\right)\right\}$, belong to the natural numbers $\mathbb{N}_{0}$.
These dual variables are the expansion indices of the power series expansion of the Boltzmann 
factor for each plaquette. The partition function in terms of the new variables is given by a 
sum of weights $Q_{\left\{ U\right\} }\left[\left\{ b\right\} \right]$. The PONs are subject
to constraints imposed by gauge symmetry, given by Eq.~(\ref{eq:vinculo}). 
  
The weights $Q$ for a configuration of PONs involve integrals over the link variables over 
the entire lattice volume whose integrands are products of powers of plaquettes. We used Monte 
Carlo simulations to solve the theory. The transition probability $p$ defining the corresponding 
Markov chain is given by ratios of the weights $Q$. The link integration is simplified to get an 
approximate analytical estimation for $p$. Specifically, the approximations consist in the 
following. In an update of $b_{\mu\nu}\left(x\right)$, the lattice is factorized into a region 
containing the plaquette $U_{\mu\nu}\left(x\right)$ and its complementary. Then, the links at the 
interface between the two regions are rotated to the identity. This allows us to evaluate 
analytically the link integrals necessary to estimate $p$. Two different types 
of updates, named local and nonlocal, are considered. In the local updates, a given PON 
$b_{\mu\nu}(x)$ is chosen randomly and a change by $\pm2${\textemdash}we have concentrated on 
SU(2) gauge theory, see Eq.~(\ref{eq:vinculo}){\textemdash}is proposed with the sign being chosen 
randomly. For the nonlocal update, a two dimensional surface is chosen randomly on the lattice and 
for each PON $b_{\mu\nu}(x)$ on the surface a change by $\pm1$ is proposed randomly. The nonlocal 
update improves the ergodicity of the algorithm as it allows to switch the occupation numbers 
from odd to even and vice-versa, an evolution which is not allowed by the local updates. We have 
not considered updates that involve changing the PONs on a cube. 

The estimations for the plaquette mean value agree very well with those obtained with a conventional 
heath bath algorithm in the weak and strong coupling limits. Deviations from heath bath estimations 
occur in the range $2.5 < \beta < 3$, but they are below than 0.1\%. In what concerns the estimation 
of the lightest SU(2) glueball mass, the simulations reported here are in good agreement with estimated 
in the literature.

We stress that the approach presented here relies on a series of approximations to get the 
transition probability $p$. 
One can speculate that the fact that the links rotated to the identity are a very small subset of 
the entire set of links $\{ U_\mu (x) \}$, their contribution to $p$ to be subleading,
at least for the quantities studied. Furthermore, given that the number of links set to identity is 
volume independent, one expects to approximate the exact value of $p$ in the limit of large volumes. 

The results reported here suggest that the inclusion of larger lattice partitions in the ``inner'' 
integral, i.e. including larger numbers of links in the neighborhood of the updated plaquette 
occupation number, to estimate the transition probability takes $p$ closer to its real value.
This can be achieved by a careful choice of the ``inner" region, i.e. the region which includes the 
lattice point where the plaquette occupation number is to be updated, and the ``outer" sublattices such 
that one is able to perform necessary group integrals after setting some of the links to the identity. 
Certainly, any progress in the evaluation of SU(N) integrals, see e.g. Ref.~\cite{Zuber:2016xme}, will 
help in improving  the estimation of $p$. Another possible approach, still to be developed, is the 
numerical evaluation of the group integrals which, hopefully, could lead to an ``exact'' estimation 
of the transition probability.

The algorithm discussed here can be generalized to SU(N) gauge groups with N > 2. Another interesting 
research topic is the inclusion of the fermionic degrees of freedom which, in principle, can be 
accommodated within the procedure described. These are research problems that we aim to address 
in the near future.

\begin{acknowledgements}
The authors thank Paulo J. Silva for the heat bath data.
This research was supported by computational resources supplied by the Center for Scientific 
Computing (NCC/GridUNESP) of the S\~ao Paulo State University (UNESP) and by the Departament of 
Physics of Coimbra University.
Work partially supported by Funda\c{c}\~ao de Amparo \`a Pesquisa do Estado de 
S\~ao Paulo (FAPESP), Grant. No. 2013/01907-0 (G.K.) and No. 2017/01142-4 (O.O.), 
Conselho Nacional de Desenvolvimento Cient\'{\i}fico e Tecnol\'ogico (CNPq), Grant No. 305894/2009-9 (G.K.) 
and No. 464898/2014-5 (G.K. and O.O.) (INCT F\'{\i}sica  Nuclear e Applica\c{c}\~oes), and  by  
Coordena\c{c}\~ao de Aperfei\c{c}oamento de Pessoal de N\'{\i}vel Superior (CAPES) (R.L.). 
\label{sec:Acknowledgments} 
\end{acknowledgements}

\appendix

\section{Group integration\label{sec:Group-integration}}

Here we will compute some group integrations that appears in the evaluation
of the weight function of the non-Abelian gauge partition function
written in the new representation. First we introduce some basic
properties of the group integration. Consider a function $f(U)$ where
$U\in\textrm{SU(N)}$, as we can see in many textbooks e.g. \cite{gattBook},
the group integration is left and right invariant

\begin{eqnarray}
\int dU\,f(U)=\int dU\,f(U\,g)=\int dU\,f(g'\,U)\label{left_right_inv}
\end{eqnarray}
where $g$ and $g'$ are arbitrary elements of the group SU(N) and
the Haar measure are also left and right invariant 
\begin{eqnarray}
dU=d(Ug)=d(g'U).
\end{eqnarray}
From these basic properties we can conclude 
\begin{eqnarray}
\int & dU & U_{a,b}=0,\label{eq:haar_prop_1}\\
\int & dU & U_{a,b}U_{c,d}^{\dagger}=\frac{1}{N}\delta_{a,d}\delta_{b,c},\label{eq:haar_prop_2}\\
\int & dU & U_{a_{1},b_{1}}\dots U_{a_{k},b_{k}}\neq0{\rm \quad\textrm{If}\,}k{\rm \,\textrm{mod}\,}\textrm{N}=0,\label{eq:haar_prop_3}
\end{eqnarray}
and these properties determine the constraint over the new degrees
of freedom discussed in Sec.~\ref{sub:Constraint-over-the}.

\subsection{Integration over the green paths}

Consider the green path, see Fig.~\ref{fig:3d_zoom-1}, containing
the links variables $U_{3a}$,$U_{3b}$, $U_{3c}$ and $U_{3d}$ that
need be integrated, this path is coupled to the central plaquette
(CP) $U_{\mu_{0}\nu_{0}}\left(x_{0}\right)= U_{1}U_{2}U_{3}U_{4}$
by the link $U_{3}$ and belong in a plane parallel to CP in the direction
$\rho$. Each link of CP is coupled to one green path, here we will
show the integration of the green path coupled to the link $U_{3}$,
the precise definition of this green path links are 
\begin{eqnarray}
U_{3a} & = & U_{\nu_{0}}(x_{0}+\hat{\nu}_{0}+\hat{\rho}-\hat{\mu}_{0}),\\
U_{3b} & = & U_{\mu_{0}}(x_{0}+2\hat{\nu}_{0}+\hat{\rho}-\hat{\mu}_{0}),\\
U_{3c} & = & U_{\nu_{0}}^{\dagger}(x_{0}+\hat{\nu}_{0}+\hat{\rho}),\\
U_{3d} & = & U_{\mu_{0}}(x_{0}+\hat{\nu}_{0}+\hat{\rho}),
\end{eqnarray}
and the integration in question is given by 
\begin{eqnarray}
P_{\mathcal{B}_{G}}[U_{3}] & = & \int\widetilde{\mathscr{D}U}_{G}\textrm{Tr}\left[U_{3}U_{3d}\right]^{b_{1}}\textrm{Tr}\left[U_{3c}U_{3d}\right]^{b_{2}}\nonumber \\
 & \times & \textrm{Tr}\left[U_{3a}U_{3b}U_{3c}\right]^{b_{3}}\mbox{Tr}\left[U_{3a}\right]^{c_{1}}\nonumber \\
 & \times & \mbox{Tr}\left[U_{3b}\right]^{c_{2}}\textrm{Tr}\left[U_{3c}\right]^{c_{3}}\textrm{Tr}\left[U_{3d}\right]^{c_{4}}.\label{eq:int_green_detailed}
\end{eqnarray}
The plaquette occupation numbers (PON) $b_{i}$ are defined as 
\begin{eqnarray}
b_{1} & = & b_{\mu_{0}\rho}(x_{0}+\hat{\nu}_{0}),\\
b_{2} & = & b_{\mu_{0}\nu_{0}}(x_{0}+\hat{\nu}_{0}+\hat{\rho}),\\
b_{3} & = & b_{\mu_{0}\nu_{0}}(x_{0}+\hat{\nu}_{0}+\hat{\rho}-\hat{\mu}_{0}),
\end{eqnarray}
and the collective powers $c_{i}$ are a sum of PONs and, using 
Eq.~\eqref{eq:total_flux}, are defined as 
\begin{eqnarray}
c_{1} & = & n_{\nu_{0}}(x_{0}+\hat{\nu}_{0}+\hat{\rho}-\hat{\mu}_{0})-b_{3},\\
c_{2} & = & n_{\mu_{0}}(x_{0}+2\hat{\nu}_{0}+\hat{\rho}-\hat{\mu}_{0})-b_{3},\\
c_{3} & = & n_{\nu_{0}}(x_{0}+\hat{\nu}_{0}+\hat{\rho})-b_{3}-b_{2},\\
c_{4} & = & n_{\mu_{0}}(x_{0}+\hat{\nu}_{0}+\hat{\rho})-b_{2}-b_{1}.
\end{eqnarray}
We use Eq.~\eqref{eq:basic_int_sol} to solve each integral
in the path. Starting the integration by the link $U_{3a}$ we have
\begin{eqnarray}
P_{\mathcal{B}_{G}}[U_{3}] & = & \sum_{q_{1}}^{\min\left(b_{3},c_{1}\right)}\Gamma_{q_{1}}^{b_{3},c_{1}}\int\widetilde{\mathscr{D}U}'_{G}K_{1},\label{eq:green_1}
\end{eqnarray}
where the coefficients $\Gamma$ are given by Eq.~\eqref{eq:coeff}.
The function $K_{1}=K_{1}\left[U_{3b},U_{3c},U_{3d};U_{3},\left\{ b'\right\} \right]$
is defined by 
\begin{eqnarray}
K_{1} & = & \textrm{Tr}\left[U_{3}U_{3d}\right]^{b_{1}}\textrm{Tr}\left[U_{3c}U_{3d}\right]^{b_{2}}\mbox{Tr}\left[U_{3b}\right]^{c_{2}}\nonumber \\
 & \times & \textrm{Tr}\left[U_{3c}\right]^{c_{3}}\textrm{Tr}\left[U_{3d}\right]^{c_{4}}\textrm{Tr}\left[U_{3b}U_{3c}\right]^{q_{1}},
\end{eqnarray}
and the measure $\widetilde{\mathscr{D}U}'_{G}$ by
\begin{equation}
\widetilde{\mathscr{D}U}'_{G}= dU_{3b}dU_{3c}dU_{3d}.
\end{equation}
Now integrating $K_{1}$ with the measure $dU_{3b}$ we find 
\begin{equation}
\int\widetilde{\mathscr{D}U}'_{G}K_{1}=\sum_{q_{2}}^{\min\left(q_{1},c_{2}\right)}\Gamma_{q_{2}}^{q_{1},c_{2}}\int\widetilde{\mathscr{D}U}''_{p1}K_{2},\label{eq:green_2}
\end{equation}
where $K_{2}=K_{2}\left[U_{3c},U_{3d};U_{3},\left\{ b'\right\} \right]$
and the measure $\widetilde{\mathscr{D}U}''_{G}$ are defined by 
\begin{eqnarray}
K_{2} & = & \textrm{Tr}\left[U_{3}U_{3d}\right]^{b_{1}}\textrm{Tr}\left[U_{3c}U_{3d}\right]^{b_{2}}\nonumber \\
 & \times & \textrm{Tr}\left[U_{3c}\right]^{c_{3}+q_{2}}\textrm{Tr}\left[U_{3d}\right]^{c_{4}},\\
\widetilde{\mathscr{D}U}''_{G} & = & dU_{3c}dU_{3d}.
\end{eqnarray}
Integrating the link $U_{3c}$ we obtain{\small{}
\begin{equation}
\int\widetilde{\mathscr{D}U}''_{G}K_{2}=\sum_{q_{3}}^{\min\left(b_{2},c_{3}+q_{2}\right)}\Gamma_{q_{3}}^{b_{2},c_{3}+q_{2}}\int dU_{3d}K_{3},\label{eq:green_3}
\end{equation}
}where $K_{3}=K_{3}\left[U_{3d};U_{3},\left\{ b'\right\} \right]$
is defined by 
\begin{eqnarray}
K_{3} & = & \textrm{Tr}\left[U_{3}U_{3d}\right]^{b_{1}}\textrm{Tr}\left[U_{3d}\right]^{c_{4}+q_{3}}.
\end{eqnarray}
Finally integrating the link $U_{3d}$ we have 
\begin{equation}
\int dU_{3d}K_{3}=\sum_{q_{4}}^{\min\left(b_{1},c_{4}+q_{3}\right)}\Gamma_{q_{4}}^{b_{1},c_{4}+q_{3}}\textrm{Tr}\left[U_{3}\right]^{q_{4}}.\label{eq:green_4}
\end{equation}
Collecting Eqs.~\eqref{eq:green_1}, \eqref{eq:green_2}, \eqref{eq:green_3} and \eqref{eq:green_4},
we have 
\begin{eqnarray}
P_{\mathcal{B}_{G}}[U_{3}] & = & \sum_{q_{1}}^{\min\left(b_{3},c_{1}\right)}\,
\sum_{q_{2}}^{\min\left(q_{1},c_{2}\right)} \, \sum_{q_{3}}^{\min\left(b_{2},c_{3}+q_{2}\right)}
\sum_{q_{4}}^{\min\left(b_{1},c_{4}+q_{3}\right)} \nonumber \\
&& \times \, 
 \Gamma_{q_{1}}^{b_{3},c_{1}}
\Gamma_{q_{2}}^{q_{1},c_{2}}
\Gamma_{q_{3}}^{b_{2},c_{3}+q_{2}}
\Gamma_{q_{4}}^{b_{1},c_{4}+q_{3}} \nonumber \\[0.25true cm]
&&\times \, \textrm{Tr}\left[U_{3}\right]^{q_{4}},
\label{eq:int_black_detailed-1}
\end{eqnarray}
i.e., a polynomial in $\textrm{Tr}\left[U_{3}\right]$. This solution
can be applied to the other green paths but the definitions of the
green path links, the PONs $b_{i}$ and the sum of PONs $c_{i}$ change
accordingly.

\subsection{Integration over the blue paths}

In the Fig.~\ref{fig:3d_zoom-2} we present the blue path coupled
to the link $U_{3}$ of CP. Like in the green path case, each link
of CP is coupled to one blue path. Here we will show only the integration
of the blue path coupled to the link $U_{3}$, the integration in
question is 
\begin{eqnarray}
P_{\mathcal{B}_{B}}[U_{3}] & = & \int\widetilde{\mathscr{D}U}_{B}\textrm{Tr}\left[\tilde{U}_{3a}\tilde{U}_{3b}\tilde{U}_{3c}\tilde{U}_{3d}\right]^{\tilde{b}_{2}}\nonumber \\
 & \times & \textrm{Tr}\left[U_{3}^{\dagger}\tilde{U}_{3e}\tilde{U}_{3d}^{\dagger}\right]^{\tilde{b}_{1}}\textrm{Tr}\left[\tilde{U}_{3a}\right]^{\tilde{c}_{1}}\textrm{Tr}\left[\tilde{U}_{3b}\right]^{\tilde{c}_{2}}\nonumber \\
 & \times & \textrm{Tr}\left[\tilde{U}_{3c}\right]^{\tilde{c}_{3}}\textrm{Tr}\left[\tilde{U}_{3d}\right]^{\tilde{c}_{4}}\textrm{Tr}\left[\tilde{U}_{3e}\right]^{\tilde{c}_{5}},\label{eq:int_blue_detailed}
\end{eqnarray}
where the definitions of the blue path links are 
\begin{eqnarray}
\tilde{U}_{3a} & = & U_{\nu_{0}}(x_{0}+2\hat{\nu}_{0}+\hat{\mu}_{0}),\\
\tilde{U}_{3b} & = & U_{\mu_{0}}^{\dagger}(x_{0}+3\hat{\nu}_{0}),\\
\tilde{U}_{3c} & = & U_{\nu_{0}}^{\dagger}(x_{0}+2\hat{\nu}_{0}),\\
\tilde{U}_{3d} & = & U_{\mu_{0}}(x_{0}+2\hat{\nu}_{0}),\\
\tilde{U}_{3e} & = & U_{\nu_{0}}(x_{0}+\hat{\nu}_{0}+\hat{\mu}_{0}).
\end{eqnarray}
The PONs $\tilde{b_{i}}$ are defined as 
\begin{eqnarray}
\tilde{b}_{1} & = & b_{\mu_{0}\nu_{0}}(x_{0}+\hat{\nu}_{0}),\\
\tilde{b}_{2} & = & b_{\mu_{0}\nu_{0}}(x_{0}+2\hat{\nu}_{0}),
\end{eqnarray}
and the collective powers $\tilde{c_{i}}$ are defined by
\begin{eqnarray}
\tilde{c}_{1} & = & n_{\nu_{0}}(x_{0}+2\hat{\nu}_{0}+\hat{\mu}_{0})-b_{2},\\
\tilde{c}_{2} & = & n_{\mu_{0}}(x_{0}+3\hat{\nu}_{0})-b_{2},\\
\tilde{c}_{3} & = & n_{\nu_{0}}(x_{0}+2\hat{\nu}_{0})-b_{2},\\
\tilde{c}_{4} & = & n_{\mu_{0}}(x_{0}+2\hat{\nu}_{0})-b_{2}-b_{1},\\
\tilde{c}_{5} & = & n_{\mu_{0}}(x_{0}+\hat{\nu}_{0}+\hat{\mu}_{0})-b_{1}.
\end{eqnarray}
In order to guarantee that we  deal with integration that looks
like Eq.~\eqref{eq:basic_int_sol}, we need start the integration
by one link of the plaquette $\tilde{U}_{3a}\tilde{U}_{3b}\tilde{U}_{3c}\tilde{U}_{3d}$,
here we start by the link $\tilde{U}_{3a}$, then
\begin{eqnarray}
P_{\mathcal{B}_{B}}[U_{3}] & = & \sum_{\tilde{q}_{1}}^{\min\left(\tilde{b}_{2},\tilde{c}_{1}\right)}
\Gamma_{\tilde{q}_{1}}^{\tilde{b}_{2},\tilde{c}_{1}}\int\widetilde{\mathscr{D}U}'_{B}\tilde{K}_{1},\label{eq:blue_1}
\end{eqnarray}
where $\tilde{K}_{1}$ and the measure $\widetilde{\mathscr{D}U}'_{B}$
are defined by 
\begin{eqnarray}
\tilde{K}_{1} & = & \textrm{Tr}\left[U_{3}^{\dagger}\tilde{U}_{3e}\tilde{U}_{3d}^{\dagger}\right]^{\tilde{b}_{1}}\textrm{Tr}\left[\tilde{U}_{3b}\tilde{U}_{3c}\tilde{U}_{3d}\right]^{\tilde{q}_{1}}\nonumber \\
 & \times & \textrm{Tr}\left[\tilde{U}_{3b}\right]^{\tilde{c}_{2}}\textrm{Tr}\left[\tilde{U}_{3c}\right]^{\tilde{c}_{3}}\textrm{Tr}\left[\tilde{U}_{3d}\right]^{\tilde{c}_{4}}\textrm{Tr}\left[\tilde{U}_{3e}\right]^{\tilde{c}_{5}},\nonumber \\
\\
\widetilde{\mathscr{D}U}'_{B} & = & d\tilde{U}_{3b}d\tilde{U}_{3c}d\tilde{U}_{3d}d\tilde{U}_{3e}.
\end{eqnarray}
Now integrating $\tilde{K}_{1}$ by the measure $d\tilde{U}_{3b}$
we have
\begin{equation}
\int\widetilde{\mathscr{D}U}'_{B}\tilde{K}_{1}=\sum_{\tilde{q}_{2}}^{\min\left(\tilde{q}_{1},\tilde{c}_{2}\right)}\Gamma_{\tilde{q}_{2}}^{\tilde{q}_{1},\tilde{c}_{2}}\int\widetilde{\mathscr{D}U}''_{B}\tilde{K}_{2},\label{eq:blue_2}
\end{equation}
where $\tilde{K}_{2}$ and the measure $\widetilde{\mathscr{D}U}''_{B}$
are defined by 
\begin{eqnarray}
\tilde{K}_{2} & = & \textrm{Tr}\left[U_{3}^{\dagger}\tilde{U}_{3e}\tilde{U}_{3d}^{\dagger}\right]^{\tilde{b}_{1}}
\textrm{Tr}\left[\tilde{U}_{3c}\right]^{\tilde{c}_{3}}\textrm{Tr}\left[\tilde{U}_{3d}\right]^{\tilde{c}_{4}}
\nonumber \\
 & \times & \textrm{Tr}\left[\tilde{U}_{3e}\right]^{\tilde{c}_{5}}\textrm{Tr}\left[\tilde{U}_{3c}\tilde{U}_{3d}
 \right]^{\tilde{q}_{2}},\\
\widetilde{\mathscr{D}U}''_{B} & = & d\tilde{U}_{3c}d\tilde{U}_{3d}d\tilde{U}_{3e}.
\end{eqnarray}
Integrating the link $\tilde{U}_{3c}$ we obtain 
\begin{equation}
\int\widetilde{\mathscr{D}U}''_{B}\tilde{K}_{2}=\sum_{\tilde{q}_{3}}^{\min\left(\tilde{q}_{2},\tilde{c}_{3}\right)}
\Gamma_{\tilde{q}_{3}}^{\tilde{q}_{2},\tilde{c}_{3}}\int\widetilde{\mathscr{D}U}'''_{B}\tilde{K}_{3},\label{eq:blue_3}
\end{equation}
where $\tilde{K}_{3}$ and the measure $\widetilde{\mathscr{D}U}'''_{B}$
are defined by 
\begin{eqnarray}
\tilde{K}_{3} & = & \textrm{Tr}\left[U_{3}^{\dagger}\tilde{U}_{3e}\tilde{U}_{3d}^{\dagger}\right]^{\tilde{b}_{1}}\textrm{Tr}\left[\tilde{U}_{3e}\right]^{\tilde{c}_{5}}\nonumber \\
& \times & \textrm{Tr}\left[\tilde{U}_{3d}\right]^{\tilde{c}_{4}+\tilde{q}_{3}},\\
\widetilde{\mathscr{D}U}'''_{B} & = & d\tilde{U}_{3d}d\tilde{U}_{3e}.
\end{eqnarray}
Now integrating the link $\tilde{U}_{3d}$ we can write
\begin{equation}
\int\widetilde{\mathscr{D}U}'''_{B}\tilde{K}_{3}=\sum_{\tilde{q}_{4}}^{\min\left(\tilde{b}_{1},\tilde{c}_{4}+\tilde{q}_{3}\right)}\Gamma_{\tilde{q}_{4}}^{\tilde{b}_{1},\tilde{c}_{4}+\tilde{q}_{3}}\int d\tilde{U}_{e}\tilde{K}_{4},\label{eq:blue_4}
\end{equation}
where $\tilde{K}_{4}$ is defined by 
\begin{eqnarray}
\tilde{K}_{4} & = & \textrm{Tr}\left[U_{3}^{\dagger}\tilde{U}_{3e}\right]^{\tilde{q}_{4}}\textrm{Tr}\left[\tilde{U}_{3e}\right]^{\tilde{c}_{5}}.
\end{eqnarray}
Finally, integrating the link $\tilde{U}_{3e}$ we have
\begin{equation}
\int d\tilde{U}_{e}\tilde{K}_{4}=\sum_{\tilde{q}_{5}}^{\min\left(\tilde{q}_{4},\tilde{c}_{5}\right)}\Gamma_{\tilde{q}_{5}}^{\tilde{q}_{4},\tilde{c}_{5}}\textrm{Tr}\left[U\right]^{\tilde{q}_{5}}.\label{eq:blue_5}
\end{equation}
Now, inserting Eqs.~\eqref{eq:blue_2},~\eqref{eq:blue_3},~\eqref{eq:blue_4}
and~\eqref{eq:blue_5} into Eq.~\eqref{eq:blue_1} we can write
\begin{eqnarray}
P_{\mathcal{B}_{B}}[U_{3}] & = & \sum_{\tilde{q}_{1}}^{\min\left(\tilde{b}_{2},\tilde{c}_{1}\right)}
\sum_{\tilde{q}_{2}}^{\min\left(\tilde{q}_{1},\tilde{c}_{2}\right)}
\sum_{\tilde{q}_{3}}^{\min\left(\tilde{q}_{2},\tilde{c}_{3}\right)} 
 \sum_{\tilde{q}_{4}}^{\min\left(\tilde{b}_{1},\tilde{c}_{4}+\tilde{q}_{3}\right)} \nonumber \\
&& \times \, \sum_{\tilde{q}_{5}}^{\min\left(\tilde{q}_{4},\tilde{c}_{5}\right)}
\Gamma_{\tilde{q}_{1}}^{\tilde{b}_{2},\tilde{c}_{1}}
\Gamma_{\tilde{q}_{2}}^{\tilde{q}_{1},\tilde{c}_{2}}
\Gamma_{\tilde{q}_{2}}^{\tilde{q}_{2},\tilde{c}_{3}}
\Gamma_{\tilde{q}_{3}}^{\tilde{b}_{1},\tilde{c}_{4}+\tilde{q}_{3}}
\nonumber \\
&& \times \, \Gamma_{\tilde{q}_{5}}^{\tilde{q}_{4},\tilde{c}_{5}} \,
\textrm{Tr}\left[U_{3}\right]^{\tilde{q}_{5}}.\label{eq:int_blue_detailed-1}
\end{eqnarray}
As in the green path case, the outcome is a polynomial. The above solution
can be used to evaluate all blue path integrations but we need to be careful,
the definitions of the blue path links, the PONs $b_{i}$ and the
sum of PONs $c_{i}$ change accordingly.

\bibliographystyle{spbasic}

\begin{thebibliography}{50}%
\makeatletter
\providecommand \@ifxundefined [1]{%
 \@ifx{#1\undefined}
}%
\providecommand \@ifnum [1]{%
 \ifnum #1\expandafter \@firstoftwo
 \else \expandafter \@secondoftwo
 \fi
}%
\providecommand \@ifx [1]{%
 \ifx #1\expandafter \@firstoftwo
 \else \expandafter \@secondoftwo
 \fi
}%
\providecommand \natexlab [1]{#1}%
\providecommand \enquote  [1]{``#1''}%
\providecommand \bibnamefont  [1]{#1}%
\providecommand \bibfnamefont [1]{#1}%
\providecommand \citenamefont [1]{#1}%
\providecommand \href@noop [0]{\@secondoftwo}%
\providecommand \href [0]{\begingroup \@sanitize@url \@href}%
\providecommand \@href[1]{\@@startlink{#1}\@@href}%
\providecommand \@@href[1]{\endgroup#1\@@endlink}%
\providecommand \@sanitize@url [0]{\catcode `\\12\catcode `\$12\catcode
  `\&12\catcode `\#12\catcode `\^12\catcode `\_12\catcode `\%12\relax}%
\providecommand \@@startlink[1]{}%
\providecommand \@@endlink[0]{}%
\providecommand \url  [0]{\begingroup\@sanitize@url \@url }%
\providecommand \@url [1]{\endgroup\@href {#1}{\urlprefix }}%
\providecommand \urlprefix  [0]{URL }%
\providecommand \Eprint [0]{\href }%
\providecommand \doibase [0]{http://dx.doi.org/}%
\providecommand \selectlanguage [0]{\@gobble}%
\providecommand \bibinfo  [0]{\@secondoftwo}%
\providecommand \bibfield  [0]{\@secondoftwo}%
\providecommand \translation [1]{[#1]}%
\providecommand \BibitemOpen [0]{}%
\providecommand \bibitemStop [0]{}%
\providecommand \bibitemNoStop [0]{.\EOS\space}%
\providecommand \EOS [0]{\spacefactor3000\relax}%
\providecommand \BibitemShut  [1]{\csname bibitem#1\endcsname}%
\let\auto@bib@innerbib\@empty
\bibitem [{\citenamefont {Gattringer}\ and\ \citenamefont
  {Lang}(2010)}]{gattBook}%
  \BibitemOpen
  \bibfield  {author} {\bibinfo {author} {\bibfnamefont {C.}~\bibnamefont
  {Gattringer}}\ and\ \bibinfo {author} {\bibfnamefont {C.~B.}\ \bibnamefont
  {Lang}},\ }\href@noop {} {\emph {\bibinfo {title} {Quantum Chromodynamics on
  the lattice: An Introductory Presentation}}}\ (\bibinfo  {publisher}
  {Springer},\ \bibinfo {year} {2010})\BibitemShut {NoStop}%
\bibitem [{\citenamefont {Splittorff}\ and\ \citenamefont
  {Svetitsky}(2007)}]{Splittorff:2007mr}%
  \BibitemOpen
  \bibfield  {author} {\bibinfo {author} {\bibfnamefont {K.}~\bibnamefont
  {Splittorff}}\ and\ \bibinfo {author} {\bibfnamefont {B.}~\bibnamefont
  {Svetitsky}},\ }\href@noop {} {\bibfield  {journal} {\bibinfo  {journal}
  {Phys. Rev.}\ }\textbf {\bibinfo {volume} {D75}},\ \bibinfo {pages} {114504}
  (\bibinfo {year} {2007})}\BibitemShut {NoStop}%
\bibitem [{\citenamefont {Splittorff}\ and\ \citenamefont
  {Verbaarschot}(2007)}]{Splittorff:2007ck}%
  \BibitemOpen
  \bibfield  {author} {\bibinfo {author} {\bibfnamefont {K.}~\bibnamefont
  {Splittorff}}\ and\ \bibinfo {author} {\bibfnamefont {J.~J.~M.}\ \bibnamefont
  {Verbaarschot}},\ }\href@noop {} {\bibfield  {journal} {\bibinfo  {journal}
  {Phys. Rev.}\ }\textbf {\bibinfo {volume} {D75}},\ \bibinfo {pages} {116003}
  (\bibinfo {year} {2007})}\BibitemShut {NoStop}%
\bibitem [{\citenamefont {Gattringer}\ and\ \citenamefont
  {Langfeld}(2016)}]{Gattringer:2016kco}%
  \BibitemOpen
  \bibfield  {author} {\bibinfo {author} {\bibfnamefont {C.}~\bibnamefont
  {Gattringer}}\ and\ \bibinfo {author} {\bibfnamefont {K.}~\bibnamefont
  {Langfeld}},\ }\href@noop {} {\bibfield  {journal} {\bibinfo  {journal} {Int.
  J. Mod. Phys.}\ }\textbf {\bibinfo {volume} {A31}},\ \bibinfo {pages}
  {1643007} (\bibinfo {year} {2016})}\BibitemShut {NoStop}%
\bibitem [{\citenamefont {Aarts}(2015)}]{Aarts:2013naa}%
  \BibitemOpen
  \bibfield  {author} {\bibinfo {author} {\bibfnamefont {G.}~\bibnamefont
  {Aarts}},\ }\href@noop {} {\bibfield  {journal} {\bibinfo  {journal}
  {Pramana}\ }\textbf {\bibinfo {volume} {84}},\ \bibinfo {pages} {787}
  (\bibinfo {year} {2015})}\BibitemShut {NoStop}%
\bibitem [{\citenamefont {de~Forcrand}\ \emph {et~al.}(2014)\citenamefont
  {de~Forcrand}, \citenamefont {Langelage}, \citenamefont {Philipsen},\ and\
  \citenamefont {Unger}}]{deForcrand:2014tha}%
  \BibitemOpen
  \bibfield  {author} {\bibinfo {author} {\bibfnamefont {P.}~\bibnamefont
  {de~Forcrand}}, \bibinfo {author} {\bibfnamefont {J.}~\bibnamefont
  {Langelage}}, \bibinfo {author} {\bibfnamefont {O.}~\bibnamefont
  {Philipsen}}, \ and\ \bibinfo {author} {\bibfnamefont {W.}~\bibnamefont
  {Unger}},\ }\href@noop {} {\bibfield  {journal} {\bibinfo  {journal} {Phys.
  Rev. Lett.}\ }\textbf {\bibinfo {volume} {113}},\ \bibinfo {pages} {152002}
  (\bibinfo {year} {2014})}\BibitemShut {NoStop}%
\bibitem [{\citenamefont {de~Forcrand}\ and\ \citenamefont
  {Fromm}(2010)}]{deForcrand:2009dh}%
  \BibitemOpen
  \bibfield  {author} {\bibinfo {author} {\bibfnamefont {P.}~\bibnamefont
  {de~Forcrand}}\ and\ \bibinfo {author} {\bibfnamefont {M.}~\bibnamefont
  {Fromm}},\ }\href@noop {} {\bibfield  {journal} {\bibinfo  {journal} {Phys.
  Rev. Lett.}\ }\textbf {\bibinfo {volume} {104}},\ \bibinfo {pages} {112005}
  (\bibinfo {year} {2010})}\BibitemShut {NoStop}%
\bibitem [{\citenamefont {Rossi}\ and\ \citenamefont
  {Wolff}(1984)}]{Rossi:1984cv}%
  \BibitemOpen
  \bibfield  {author} {\bibinfo {author} {\bibfnamefont {P.}~\bibnamefont
  {Rossi}}\ and\ \bibinfo {author} {\bibfnamefont {U.}~\bibnamefont {Wolff}},\
  }\href@noop {} {\bibfield  {journal} {\bibinfo  {journal} {Nucl. Phys.}\
  }\textbf {\bibinfo {volume} {B248}},\ \bibinfo {pages} {105} (\bibinfo {year}
  {1984})}\BibitemShut {NoStop}%
\bibitem [{\citenamefont {Prokof'ev}\ and\ \citenamefont
  {Svistunov}(2001)}]{Prokofev:2001ddj}%
  \BibitemOpen
  \bibfield  {author} {\bibinfo {author} {\bibfnamefont {N.}~\bibnamefont
  {Prokof'ev}}\ and\ \bibinfo {author} {\bibfnamefont {B.}~\bibnamefont
  {Svistunov}},\ }\href@noop {} {\bibfield  {journal} {\bibinfo  {journal}
  {Phys. Rev. Lett.}\ }\textbf {\bibinfo {volume} {87}},\ \bibinfo {pages}
  {160601} (\bibinfo {year} {2001})}\BibitemShut {NoStop}%
\bibitem [{\citenamefont {DeGrand}\ and\ \citenamefont
  {DeTar}(1983)}]{DeGrand:1983fk}%
  \BibitemOpen
  \bibfield  {author} {\bibinfo {author} {\bibfnamefont {T.~A.}\ \bibnamefont
  {DeGrand}}\ and\ \bibinfo {author} {\bibfnamefont {C.~E.}\ \bibnamefont
  {DeTar}},\ }\href@noop {} {\bibfield  {journal} {\bibinfo  {journal} {Nucl.
  Phys.}\ }\textbf {\bibinfo {volume} {B225}},\ \bibinfo {pages} {590}
  (\bibinfo {year} {1983})}\BibitemShut {NoStop}%
\bibitem [{\citenamefont {Delgado~Mercado}\ \emph {et~al.}(2011)\citenamefont
  {Delgado~Mercado}, \citenamefont {Evertz},\ and\ \citenamefont
  {Gattringer}}]{Mercado:2011ua}%
  \BibitemOpen
  \bibfield  {author} {\bibinfo {author} {\bibfnamefont {Y.D.}~\bibnamefont
  {Mercado}}, \bibinfo {author} {\bibfnamefont {H.~G.}\ \bibnamefont
  {Evertz}}, \ and\ \bibinfo {author} {\bibfnamefont {C.}~\bibnamefont
  {Gattringer}},\ }\href@noop {} {\bibfield  {journal} {\bibinfo  {journal}
  {Phys. Rev. Lett.}\ }\textbf {\bibinfo {volume} {106}},\ \bibinfo {pages}
  {222001} (\bibinfo {year} {2011})}\BibitemShut {NoStop}%
\bibitem [{\citenamefont {Wolff}(2010{\natexlab{a}})}]{Wolff:2009kp}%
  \BibitemOpen
  \bibfield  {author} {\bibinfo {author} {\bibfnamefont {U.}~\bibnamefont
  {Wolff}},\ }\href@noop {} {\bibfield  {journal} {\bibinfo  {journal} {Nucl.
  Phys.}\ }\textbf {\bibinfo {volume} {B824}},\ \bibinfo {pages} {254}
  (\bibinfo {year} {2010}{\natexlab{a}})}\BibitemShut {NoStop}%
\bibitem [{\citenamefont {Wolff}(2010{\natexlab{b}})}]{Wolff:2010qz}%
  \BibitemOpen
  \bibfield  {author} {\bibinfo {author} {\bibfnamefont {U.}~\bibnamefont
  {Wolff}},\ }\href@noop {} {\bibfield  {journal} {\bibinfo  {journal} {Nucl.
  Phys.}\ }\textbf {\bibinfo {volume} {B832}},\ \bibinfo {pages} {520}
  (\bibinfo {year} {2010}{\natexlab{b}})}\BibitemShut {NoStop}%
\bibitem [{\citenamefont {Bruckmann}\ \emph {et~al.}(2015)\citenamefont
  {Bruckmann}, \citenamefont {Gattringer}, \citenamefont {Kloiber},\ and\
  \citenamefont {Sulejmanpasic}}]{Bruckmann:2015sua}%
  \BibitemOpen
  \bibfield  {author} {\bibinfo {author} {\bibfnamefont {F.}~\bibnamefont
  {Bruckmann}}, \bibinfo {author} {\bibfnamefont {C.}~\bibnamefont
  {Gattringer}}, \bibinfo {author} {\bibfnamefont {T.}~\bibnamefont {Kloiber}},
  \ and\ \bibinfo {author} {\bibfnamefont {T.}~\bibnamefont {Sulejmanpasic}},\
  }\href@noop {} {\bibfield  {journal} {\bibinfo  {journal} {Phys. Lett.}\
  }\textbf {\bibinfo {volume} {B749}},\ \bibinfo {pages} {495} (\bibinfo {year}
  {2015})}\BibitemShut {NoStop}%
\bibitem [{\citenamefont {Castro~Neto}\ \emph {et~al.}(2009)\citenamefont
  {Castro~Neto}, \citenamefont {Guinea}, \citenamefont {Peres}, \citenamefont
  {Novoselov},\ and\ \citenamefont {Geim}}]{CastroNeto:2009zz}%
  \BibitemOpen
  \bibfield  {author} {\bibinfo {author} {\bibfnamefont {A.~H.}\ \bibnamefont
  {Castro~Neto}}, \bibinfo {author} {\bibfnamefont {F.}~\bibnamefont {Guinea}},
  \bibinfo {author} {\bibfnamefont {N.~M.~R.}\ \bibnamefont {Peres}}, \bibinfo
  {author} {\bibfnamefont {K.~S.}\ \bibnamefont {Novoselov}}, \ and\ \bibinfo
  {author} {\bibfnamefont {A.~K.}\ \bibnamefont {Geim}},\ }\href@noop {}
  {\bibfield  {journal} {\bibinfo  {journal} {Rev. Mod. Phys.}\ }\textbf
  {\bibinfo {volume} {81}},\ \bibinfo {pages} {109} (\bibinfo {year}
  {2009})}\BibitemShut {NoStop}%
\bibitem [{\citenamefont {Ayyar}\ and\ \citenamefont
  {Chandrasekharan}(2015)}]{Ayyar:2014eua}%
  \BibitemOpen
  \bibfield  {author} {\bibinfo {author} {\bibfnamefont {V.}~\bibnamefont
  {Ayyar}}\ and\ \bibinfo {author} {\bibfnamefont {S.}~\bibnamefont
  {Chandrasekharan}},\ }\href@noop {} {\bibfield  {journal} {\bibinfo
  {journal} {Phys. Rev.}\ }\textbf {\bibinfo {volume} {D91}},\ \bibinfo {pages}
  {065035} (\bibinfo {year} {2015})}\BibitemShut {NoStop}%
\bibitem [{\citenamefont {Chandrasekharan}(2013)}]{Chandrasekharan:2013rpa}%
  \BibitemOpen
  \bibfield  {author} {\bibinfo {author} {\bibfnamefont {S.}~\bibnamefont
  {Chandrasekharan}},\ }\href@noop {} {\bibfield  {journal} {\bibinfo
  {journal} {Eur. Phys. J.}\ }\textbf {\bibinfo {volume} {A49}},\ \bibinfo
  {pages} {90} (\bibinfo {year} {2013})}\BibitemShut {NoStop}%
\bibitem [{\citenamefont {Chandrasekharan}\ and\ \citenamefont
  {Li}(2013)}]{Chandrasekharan:2013aya}%
  \BibitemOpen
  \bibfield  {author} {\bibinfo {author} {\bibfnamefont {S.}~\bibnamefont
  {Chandrasekharan}}\ and\ \bibinfo {author} {\bibfnamefont {A.}~\bibnamefont
  {Li}},\ }\href@noop {} {\bibfield  {journal} {\bibinfo  {journal} {Phys.
  Rev.}\ }\textbf {\bibinfo {volume} {D88}},\ \bibinfo {pages} {021701}
  (\bibinfo {year} {2013})}\BibitemShut {NoStop}%
\bibitem [{\citenamefont {Chandrasekharan}\ and\ \citenamefont
  {Li}(2012)}]{Chandrasekharan:2011mn}%
  \BibitemOpen
  \bibfield  {author} {\bibinfo {author} {\bibfnamefont {S.}~\bibnamefont
  {Chandrasekharan}}\ and\ \bibinfo {author} {\bibfnamefont {A.}~\bibnamefont
  {Li}},\ }\href@noop {} {\bibfield  {journal} {\bibinfo  {journal} {Phys. Rev.
  Lett.}\ }\textbf {\bibinfo {volume} {108}},\ \bibinfo {pages} {140404}
  (\bibinfo {year} {2012})}\BibitemShut {NoStop}%
\bibitem [{\citenamefont {Fromm}\ \emph {et~al.}(2012)\citenamefont {Fromm},
  \citenamefont {Langelage}, \citenamefont {Lottini},\ and\ \citenamefont
  {Philipsen}}]{Fromm:2011qi}%
  \BibitemOpen
  \bibfield  {author} {\bibinfo {author} {\bibfnamefont {M.}~\bibnamefont
  {Fromm}}, \bibinfo {author} {\bibfnamefont {J.}~\bibnamefont {Langelage}},
  \bibinfo {author} {\bibfnamefont {S.}~\bibnamefont {Lottini}}, \ and\
  \bibinfo {author} {\bibfnamefont {O.}~\bibnamefont {Philipsen}},\ }\href@noop
  {} {\bibfield  {journal} {\bibinfo  {journal} {JHEP}\ }\textbf {\bibinfo
  {volume} {01}},\ \bibinfo {pages} {042} (\bibinfo {year} {2012})}\BibitemShut
  {NoStop}%
\bibitem [{\citenamefont {Gattringer}\ and\ \citenamefont
  {Kloiber}(2013{\natexlab{a}})}]{Gattringer:2012df}%
  \BibitemOpen
  \bibfield  {author} {\bibinfo {author} {\bibfnamefont {C.}~\bibnamefont
  {Gattringer}}\ and\ \bibinfo {author} {\bibfnamefont {T.}~\bibnamefont
  {Kloiber}},\ }\href@noop {} {\bibfield  {journal} {\bibinfo  {journal} {Nucl.
  Phys.}\ }\textbf {\bibinfo {volume} {B869}},\ \bibinfo {pages} {56} (\bibinfo
  {year} {2013}{\natexlab{a}})}\BibitemShut {NoStop}%
\bibitem [{\citenamefont {Korzec}\ \emph {et~al.}(2011)\citenamefont {Korzec},
  \citenamefont {Vierhaus},\ and\ \citenamefont {Wolff}}]{Korzec:2011gh}%
  \BibitemOpen
  \bibfield  {author} {\bibinfo {author} {\bibfnamefont {T.}~\bibnamefont
  {Korzec}}, \bibinfo {author} {\bibfnamefont {I.}~\bibnamefont {Vierhaus}}, \
  and\ \bibinfo {author} {\bibfnamefont {U.}~\bibnamefont {Wolff}},\
  }\href@noop {} {\bibfield  {journal} {\bibinfo  {journal} {Comput. Phys.
  Commun.}\ }\textbf {\bibinfo {volume} {182}},\ \bibinfo {pages} {1477}
  (\bibinfo {year} {2011})}\BibitemShut {NoStop}%
\bibitem [{\citenamefont {Gattringer}\ and\ \citenamefont
  {Kloiber}(2013{\natexlab{b}})}]{Gattringer:2012ap}%
  \BibitemOpen
  \bibfield  {author} {\bibinfo {author} {\bibfnamefont {C.}~\bibnamefont
  {Gattringer}}\ and\ \bibinfo {author} {\bibfnamefont {T.}~\bibnamefont
  {Kloiber}},\ }\href@noop {} {\bibfield  {journal} {\bibinfo  {journal} {Phys.
  Lett.}\ }\textbf {\bibinfo {volume} {B720}},\ \bibinfo {pages} {210}
  (\bibinfo {year} {2013}{\natexlab{b}})}\BibitemShut {NoStop}%
\bibitem [{\citenamefont {Endres}(2007)}]{Endres:2006xu}%
  \BibitemOpen
  \bibfield  {author} {\bibinfo {author} {\bibfnamefont {M.~G.}\ \bibnamefont
  {Endres}},\ }\href@noop {} {\bibfield  {journal} {\bibinfo  {journal} {Phys.
  Rev.}\ }\textbf {\bibinfo {volume} {D75}},\ \bibinfo {pages} {065012}
  (\bibinfo {year} {2007})}\BibitemShut {NoStop}%
\bibitem [{\citenamefont {Rindlisbacher}\ \emph {et~al.}(2016)\citenamefont
  {Rindlisbacher}, \citenamefont {{\AA}kerlund},\ and\ \citenamefont
  {de~Forcrand}}]{Rindlisbacher:2016zht}%
  \BibitemOpen
  \bibfield  {author} {\bibinfo {author} {\bibfnamefont {T.}~\bibnamefont
  {Rindlisbacher}}, \bibinfo {author} {\bibfnamefont {O.}~\bibnamefont
  {{\AA}kerlund}}, \ and\ \bibinfo {author} {\bibfnamefont {P.}~\bibnamefont
  {de~Forcrand}},\ }\href@noop {} {\bibfield  {journal} {\bibinfo  {journal}
  {Nucl. Phys.}\ }\textbf {\bibinfo {volume} {B909}},\ \bibinfo {pages} {542}
  (\bibinfo {year} {2016})}\BibitemShut {NoStop}%
\bibitem [{\citenamefont {Panero}(2005)}]{Panero:2005iu}%
  \BibitemOpen
  \bibfield  {author} {\bibinfo {author} {\bibfnamefont {M.}~\bibnamefont
  {Panero}},\ }\href@noop {} {\bibfield  {journal} {\bibinfo  {journal} {JHEP}\
  }\textbf {\bibinfo {volume} {05}},\ \bibinfo {pages} {066} (\bibinfo {year}
  {2005})}\BibitemShut {NoStop}%
\bibitem [{\citenamefont {Azcoiti}\ \emph {et~al.}(2009)\citenamefont
  {Azcoiti}, \citenamefont {Follana}, \citenamefont {Vaquero},\ and\
  \citenamefont {Di~Carlo}}]{Azcoiti:2009md}%
  \BibitemOpen
  \bibfield  {author} {\bibinfo {author} {\bibfnamefont {V.}~\bibnamefont
  {Azcoiti}}, \bibinfo {author} {\bibfnamefont {E.}~\bibnamefont {Follana}},
  \bibinfo {author} {\bibfnamefont {A.}~\bibnamefont {Vaquero}}, \ and\
  \bibinfo {author} {\bibfnamefont {G.}~\bibnamefont {Di~Carlo}},\ }\href@noop
  {} {\bibfield  {journal} {\bibinfo  {journal} {JHEP}\ }\textbf {\bibinfo
  {volume} {08}},\ \bibinfo {pages} {008} (\bibinfo {year} {2009})}\BibitemShut
  {NoStop}%
\bibitem [{\citenamefont {Delgado~Mercado}\ \emph
  {et~al.}(2013{\natexlab{a}})\citenamefont {Delgado~Mercado}, \citenamefont
  {Gattringer},\ and\ \citenamefont {Schmidt}}]{Mercado:2013ola}%
  \BibitemOpen
  \bibfield  {author} {\bibinfo {author} {\bibfnamefont {Y.}~\bibnamefont
  {Delgado~Mercado}}, \bibinfo {author} {\bibfnamefont {C.}~\bibnamefont
  {Gattringer}}, \ and\ \bibinfo {author} {\bibfnamefont {A.}~\bibnamefont
  {Schmidt}},\ }\href@noop {} {\bibfield  {journal} {\bibinfo  {journal} {Phys.
  Rev. Lett.}\ }\textbf {\bibinfo {volume} {111}},\ \bibinfo {pages} {141601}
  (\bibinfo {year} {2013}{\natexlab{a}})}\BibitemShut {NoStop}%
\bibitem [{\citenamefont {Delgado~Mercado}\ \emph
  {et~al.}(2013{\natexlab{b}})\citenamefont {Delgado~Mercado}, \citenamefont
  {Gattringer},\ and\ \citenamefont {Schmidt}}]{Mercado:2013yta}%
  \BibitemOpen
  \bibfield  {author} {\bibinfo {author} {\bibfnamefont {Y.}~\bibnamefont
  {Delgado~Mercado}}, \bibinfo {author} {\bibfnamefont {C.}~\bibnamefont
  {Gattringer}}, \ and\ \bibinfo {author} {\bibfnamefont {A.}~\bibnamefont
  {Schmidt}},\ }\href@noop {} {\bibfield  {journal} {\bibinfo  {journal}
  {Comput. Phys. Commun.}\ }\textbf {\bibinfo {volume} {184}},\ \bibinfo
  {pages} {1535} (\bibinfo {year} {2013}{\natexlab{b}})}\BibitemShut {NoStop}%
\bibitem [{\citenamefont {Adams}\ and\ \citenamefont
  {Chandrasekharan}(2003)}]{Adams:2003cca}%
  \BibitemOpen
  \bibfield  {author} {\bibinfo {author} {\bibfnamefont {D.~H.}\ \bibnamefont
  {Adams}}\ and\ \bibinfo {author} {\bibfnamefont {S.}~\bibnamefont
  {Chandrasekharan}},\ }\href@noop {} {\bibfield  {journal} {\bibinfo
  {journal} {Nucl. Phys.}\ }\textbf {\bibinfo {volume} {B662}},\ \bibinfo
  {pages} {220} (\bibinfo {year} {2003})}\BibitemShut {NoStop}%
\bibitem [{\citenamefont {Vairinhos}\ and\ \citenamefont
  {Forcrand}(2014)}]{vairinhos}%
  \BibitemOpen
  \bibfield  {author} {\bibinfo {author} {\bibfnamefont {H.}~\bibnamefont
  {Vairinhos}}\ and\ \bibinfo {author} {\bibfnamefont {P.}~\bibnamefont
  {Forcrand}},\ }\href@noop {} {\bibfield  {journal} {\bibinfo  {journal}
  {Journal of High Energy Physics}\ }\textbf {\bibinfo {volume} {2014}},\ \bibinfo {pages}
  {38 } (\bibinfo {year} {2014})}\BibitemShut {NoStop}%
%
 \bibitem [{\citenamefont {Hubbard}(1959)}]{hubbard}%
  \BibitemOpen
  \bibfield  {author} {\bibinfo {author} {\bibfnamefont {J.}~\bibnamefont
  {Hubbard}},\ }\href@noop {} {\bibfield  {journal} {\bibinfo  {journal}
  {Phys. Rev. Lett.}\ }\textbf {\bibinfo {volume} {3}},\ \bibinfo {pages}
  {77} (\bibinfo {year} {1959})}\BibitemShut {NoStop}%
%
\bibitem{Marchis:2017oqi}
  C.~Marchis and C.~Gattringer,
  Phys.\ Rev.\ D {\bf 97}, 034508 (2018). 
%
\bibitem [{\citenamefont {Gattringer}\ and\ \citenamefont
  {Marchis}(2017)}]{Gattringer:2011gq}%
  \BibitemOpen
  \bibfield  {author} {\bibinfo {author} {\bibfnamefont {C.}~\bibnamefont
  {Gattringer}}\ and\ \bibinfo {author} {\bibfnamefont {C.}~\bibnamefont
  {Marchis}},\ }\href@noop {} {\bibfield  {journal} {\bibinfo  {journal}
  {Nuclear Physics B}\ }\textbf {\bibinfo {volume} {916}},\ \bibinfo {pages}
  {627 } (\bibinfo {year} {2017})}\BibitemShut {NoStop}%
%
\bibitem{Marchis:2016cpe} 
  C.~Marchis and C.~Gattringer,
  PoS LATTICE {\bf 2016}, 034 (2016).
\bibitem [{\citenamefont {Chen}\ \emph {et~al.}(2006)\citenamefont {Chen} \emph
  {et~al.}}]{Chen:2005mg}%
  \BibitemOpen
  \bibfield  {author} { \bibinfo {author} {\bibfnamefont {Y.} \bibfnamefont {Chen}}, 
  \bibinfo {author} {\bibfnamefont {A.} \bibfnamefont {Alexandru}}, 
  \bibinfo {author} {\bibfnamefont {S.J.} \bibfnamefont {Dong}}, 
  \bibinfo {author} {\bibfnamefont {T.} \bibfnamefont {Draper}}, 
  \bibinfo {author} {\bibfnamefont {I.} \bibfnamefont {Horvath}}, 
  \bibinfo {author} {\bibfnamefont {F.X.} \bibfnamefont {Lee}}, 
  \bibinfo {author} {\bibfnamefont {K.F.} \bibfnamefont {Liu}}, 
  \bibinfo {author} {\bibfnamefont {N.} \bibfnamefont {Mathur}}, 
  \bibinfo {author} {\bibfnamefont {C.} \bibfnamefont {Morningstar}}, 
  \bibinfo {author} {\bibfnamefont {M.} \bibfnamefont {Peardon}}, 
  \bibinfo {author} {\bibfnamefont {S.} \bibfnamefont {Tamhankar}}, 
  \bibinfo {author} {\bibfnamefont {B.L.} \bibfnamefont {Young}}, 
  and \bibinfo {author} {\bibfnamefont {J.B.} \bibfnamefont {Zhang}} }
  \href@noop {} {\bibfield  {journal} {\bibinfo
  {journal} {Phys. Rev.}\ }\textbf {\bibinfo {volume} {D73}},\ \bibinfo {pages}
  {014516} (\bibinfo {year} {2006})}\BibitemShut {NoStop}%
\bibitem [{\citenamefont {Lucini}\ \emph {et~al.}(2004)\citenamefont {Lucini},
  \citenamefont {Teper},\ and\ \citenamefont {Wenger}}]{Lucini:2004my}%
  \BibitemOpen
  \bibfield  {author} {\bibinfo {author} {\bibfnamefont {B.}~\bibnamefont
  {Lucini}}, \bibinfo {author} {\bibfnamefont {M.}~\bibnamefont {Teper}}, \
  and\ \bibinfo {author} {\bibfnamefont {U.}~\bibnamefont {Wenger}},\
  }\href@noop {} {\bibfield  {journal} {\bibinfo  {journal} {JHEP}\ }\textbf
  {\bibinfo {volume} {06}},\ \bibinfo {pages} {012} (\bibinfo {year}
  {2004})}\BibitemShut {NoStop}%
\bibitem [{\citenamefont {Albanese}\ \emph {et~al.}(1987)\citenamefont
  {Albanese} \emph {et~al.}}]{Albanese:1987ds}%
  \BibitemOpen
  \bibfield  {author} {\bibinfo {author} {\bibfnamefont {M.}~\bibnamefont
  {Albanese}} \emph {et~al.} (\bibinfo {collaboration} {APE}),\ }\href@noop {}
  {\bibfield  {journal} {\bibinfo  {journal} {Phys. Lett.}\ }\textbf {\bibinfo
  {volume} {B192}},\ \bibinfo {pages} {163} (\bibinfo {year}
  {1987})}\BibitemShut {NoStop}%
\bibitem [{\citenamefont {Teper}(1987)}]{Teper:1987wt}%
  \BibitemOpen
  \bibfield  {author} {\bibinfo {author} {\bibfnamefont {M.}~\bibnamefont
  {Teper}},\ }\href@noop {} {\bibfield  {journal} {\bibinfo  {journal} {Phys.
  Lett.}\ }\textbf {\bibinfo {volume} {B183}},\ \bibinfo {pages} {345}
  (\bibinfo {year} {1987})}\BibitemShut {NoStop}%
\bibitem [{\citenamefont {Berg}\ and\ \citenamefont
  {Billoire}(1983)}]{Berg:1982kp}%
  \BibitemOpen
  \bibfield  {author} {\bibinfo {author} {\bibfnamefont {B.}~\bibnamefont
  {Berg}}\ and\ \bibinfo {author} {\bibfnamefont {A.}~\bibnamefont
  {Billoire}},\ }\href@noop {} {\bibfield  {journal} {\bibinfo  {journal}
  {Nucl. Phys.}\ }\textbf {\bibinfo {volume} {B221}},\ \bibinfo {pages} {109}
  (\bibinfo {year} {1983})}\BibitemShut {NoStop}%
\bibitem [{\citenamefont {Falcioni}\ \emph {et~al.}(1983)\citenamefont
  {Falcioni}, \citenamefont {Marinari}, \citenamefont {Paciello}, \citenamefont
  {Parisi}, \citenamefont {Taglienti},\ and\ \citenamefont
  {Zhang}}]{Falcioni:1982ja}%
  \BibitemOpen
  \bibfield  {author} {\bibinfo {author} {\bibfnamefont {M.}~\bibnamefont
  {Falcioni}}, \bibinfo {author} {\bibfnamefont {E.}~\bibnamefont {Marinari}},
  \bibinfo {author} {\bibfnamefont {M.~L.}\ \bibnamefont {Paciello}}, \bibinfo
  {author} {\bibfnamefont {G.}~\bibnamefont {Parisi}}, \bibinfo {author}
  {\bibfnamefont {B.}~\bibnamefont {Taglienti}}, \ and\ \bibinfo {author}
  {\bibfnamefont {Y.-c.}\ \bibnamefont {Zhang}},\ }\href@noop {} {\bibfield
  {journal} {\bibinfo  {journal} {Nucl. Phys.}\ }\textbf {\bibinfo {volume}
  {B215}},\ \bibinfo {pages} {265} (\bibinfo {year} {1983})}\BibitemShut
  {NoStop}%
\bibitem [{\citenamefont {Berg}(1980)}]{Berg:1980gz}%
  \BibitemOpen
  \bibfield  {author} {\bibinfo {author} {\bibfnamefont {B.}~\bibnamefont
  {Berg}},\ }\href@noop {} {\bibfield  {journal} {\bibinfo  {journal} {Phys.
  Lett.}\ }\textbf {\bibinfo {volume} {B97}},\ \bibinfo {pages} {401} (\bibinfo
  {year} {1980})}\BibitemShut {NoStop}%
\bibitem [{\citenamefont {Brünner}\ and\ \citenamefont
  {Rebhan}(2015)}]{Brunner:2015yha}%
  \BibitemOpen
  \bibfield  {author} {\bibinfo {author} {\bibfnamefont {F.}~\bibnamefont
  {Brunner}}\ and\ \bibinfo {author} {\bibfnamefont {A.}~\bibnamefont
  {Rebhan}},\ }\href@noop {} {\bibfield  {journal} {\bibinfo  {journal} {Phys.
  Rev. Lett.}\ }\textbf {\bibinfo {volume} {115}},\ \bibinfo {pages} {131601}
  (\bibinfo {year} {2015})}\BibitemShut {NoStop}%
\bibitem [{\citenamefont {Wilson}(1974)}]{Wilson:1974sk}%
  \BibitemOpen
  \bibfield  {author} {\bibinfo {author} {\bibfnamefont {K.~G.}\ \bibnamefont
  {Wilson}},\ }\href@noop {} {\bibfield  {journal} {\bibinfo  {journal} {Phys.
  Rev.}\ }\textbf {\bibinfo {volume} {D10}},\ \bibinfo {pages} {2445} (\bibinfo
  {year} {1974})}\BibitemShut {NoStop}%
\bibitem [{\citenamefont {Newman}\ and\ \citenamefont
  {Barkema}(1999)}]{Newman1999}%
  \BibitemOpen
  \bibfield  {author} {\bibinfo {author} {\bibfnamefont {M.}~\bibnamefont
  {Newman}}\ and\ \bibinfo {author} {\bibfnamefont {G.}~\bibnamefont
  {Barkema}},\ }\href@noop {} {\emph {\bibinfo {title} {Monte Carlo Methods in
  Statistical Physics}}}\ (\bibinfo  {publisher} {Clarendon Press},\ \bibinfo
  {year} {1999})\BibitemShut {NoStop}%
\bibitem [{\citenamefont {Creutz}(1985)}]{creutz}%
  \BibitemOpen
  \bibfield  {author} {\bibinfo {author} {\bibfnamefont {M.~J.}\ \bibnamefont
  {Creutz}},\ }\href@noop {} {\emph {\bibinfo {title} {Quarks, Gluons and
  Lattices}}}\ (\bibinfo  {publisher} {Cambridge University Press},\ \bibinfo
  {year} {1985})\BibitemShut {NoStop}%
\bibitem [{\citenamefont {Eriksson}\ \emph {et~al.}(1981)\citenamefont
  {Eriksson}, \citenamefont {Svartholm},\ and\ \citenamefont
  {Skagerstam}}]{Eriksson:1980rq}%
  \BibitemOpen
  \bibfield  {author} {\bibinfo {author} {\bibfnamefont {K.~E.}\ \bibnamefont
  {Eriksson}}, \bibinfo {author} {\bibfnamefont {N.}~\bibnamefont {Svartholm}},
  \ and\ \bibinfo {author} {\bibfnamefont {B.~S.}\ \bibnamefont {Skagerstam}},\
  }\href@noop {} {\bibfield  {journal} {\bibinfo  {journal} {J. Math. Phys.}\
  }\textbf {\bibinfo {volume} {22}},\ \bibinfo {pages} {2276} (\bibinfo {year}
  {1981})}\BibitemShut {NoStop}%
\bibitem [{\citenamefont {Edwards}\ and\ \citenamefont
  {Joo}(2005)}]{Edwards:2004sx}%
  \BibitemOpen
  \bibfield  {author} {\bibinfo {author} {\bibfnamefont {R.~G.}\ \bibnamefont
  {Edwards}}\ and\ \bibinfo {author} {\bibfnamefont {B.}~\bibnamefont {Joo}}
  (\bibinfo {collaboration} {SciDAC, LHPC, UKQCD}),\ }\href@noop {} {\bibfield
  {journal} {\bibinfo  {journal} {Nucl. Phys. Proc. Suppl.}\ }\textbf {\bibinfo
  {volume} {140}},\ \bibinfo {pages} {832} (\bibinfo {year}
  {2005})}\BibitemShut {NoStop}%
\bibitem [{\citenamefont {Bloch}\ \emph {et~al.}(2004)\citenamefont {Bloch},
  \citenamefont {Cucchieri}, \citenamefont {Langfeld},\ and\ \citenamefont
  {Mendes}}]{Bloch:2003sk}%
  \BibitemOpen
  \bibfield  {author} {\bibinfo {author} {\bibfnamefont {J.~C.~R.}\
  \bibnamefont {Bloch}}, \bibinfo {author} {\bibfnamefont {A.}~\bibnamefont
  {Cucchieri}}, \bibinfo {author} {\bibfnamefont {K.}~\bibnamefont {Langfeld}},
  \ and\ \bibinfo {author} {\bibfnamefont {T.}~\bibnamefont {Mendes}},\
  }\href@noop {} {\bibfield  {journal} {\bibinfo  {journal} {Nucl. Phys.}\
  }\textbf {\bibinfo {volume} {B687}},\ \bibinfo {pages} {76} (\bibinfo {year}
  {2004})}\BibitemShut {NoStop}%
\bibitem [{\citenamefont {Zuber}(2017)}]{Zuber:2016xme}%
  \BibitemOpen
  \bibfield  {author} {\bibinfo {author} {\bibfnamefont {J.-B.}\ \bibnamefont
  {Zuber}},\ }\href@noop {} {\bibfield  {journal} {\bibinfo  {journal} {J.
  Phys.}\ }\textbf {\bibinfo {volume} {A50}},\ \bibinfo {pages} {015203}
  (\bibinfo {year} {2017})}\BibitemShut {NoStop}%
\end{thebibliography}

\end{document}